\documentclass{aastex631}
\usepackage{lipsum}
\usepackage{pgfplots}
\pgfplotsset{compat=1.18} 

\usepackage{tikz}
  \usetikzlibrary{positioning}
  \usetikzlibrary{tikzmark} 
  \usetikzlibrary{arrows}   
  \usetikzlibrary{calc}     
  \tikzstyle{every picture}+=[remember picture]

\usepackage{CJKutf8}
\newcommand{\cntext}[1]{\begin{CJK}{UTF8}{gbsn}{\rm #1}\end{CJK}\kern-1ex}

\graphicspath{{./}{figures/}}

\begin{document}

\title{Total Power and Low-energy Cut-off Time Evolution of Solar Flare Accelerated Electrons Using X-Ray Observations and Warm-Target Model}


\correspondingauthor{Debesh Bhattacharjee}
\email{debesh.bhattacharjee@glasgow.ac.uk}
\author[0000-0002-2651-5120]{Debesh Bhattacharjee}
\affiliation{School of Physics \& Astronomy, University of Glasgow, G12 8QQ, Glasgow, UK}
\author[0000-0002-8078-0902]{Eduard P. Kontar}
\affiliation{School of Physics \& Astronomy, University of Glasgow, G12 8QQ, Glasgow, UK}
\author[0000-0002-5431-545X]{Yingjie Luo(\cntext{骆英杰})} \affiliation{School of Physics
\& Astronomy, University of Glasgow, G12 8QQ, Glasgow, UK}

\begin{abstract}
A primary characteristic of solar flares is the efficient acceleration
of electrons to nonthermal deka-keV energies. While hard X-Ray (HXR)
observation of bremsstrahlung emission serves as the key diagnostic of
these electrons. In this study, we investigate the time evolution of
flare-accelerated electrons using the warm-target model. 
This model, unlike the commonly used cold-target model, 
can determine the low-energy cut-off in the nonthermal electron distribution, 
so that the energetics of nonthermal electrons can be deduced more accurately.  
Here, we examine the time-evolution of nonthermal electrons in flares
well-observed by the RHESSI and the Solar Orbiter (SolO, using the STIX
instrument) spacecrafts. Using spectroscopic and imaging HXR
observations, the time evolution of the low-energy cut-off of the
accelerated electron distribution, the total power of nonthermal
electrons, total rate of nonthermal electrons, and excess thermal
emission measure from the nonthermal electrons, are investigated. We
find that the time profile of the low-energy cut-off of the accelerated
electron distribution shows a high-low-high trend around the HXR bursts
of flares, while the time evolution of the total rate of injected
electrons shows a low-high-low behavior. Although the total power of
nonthermal electrons is sensitive to the cut-off energy, the temporal
variation of the flare power follows the temporal variation of the
acceleration rate. We further find that the highest contribution of the
excess thermal emission measure coming from thermalization of injected
electrons takes place around the hard X-ray peak.
\end{abstract}


\section{Introduction} \label{sec:intro}

Solar flares are localized magnetic explosions on the Sun releasing up
to $\approx 10^{33}$ erg of coronal magnetic energy on timescales
of a few tens of minutes
\citep{2012ApJ...759...71E,2017ApJ...836...17A}. They can be observed
across a wide range of electromagnetic spectra --- from radio
emissions to gamma rays \citep[see, e.g.][as recent
reviews]{2011Lyndsay,2011Holman,2011Kontar,2017Benz}. Solar flares are
observed to be the primary event on the Sun for the acceleration of
electrons to tens of keVs with high efficiency \citep{2011Holman}.
Magnetic reconnection has been well documented to play a crucial role in
governing the energetics of a flare \citep{2011ShibataLRSP,1996Shibata}.
Large portion of the energy, which is stored in the magnetic fields and
released during the reconnection process, is converted into particle
kinetic energy via acceleration \citep{2017ApJ...836...17A}. These
electrons then produce X-ray signatures via the electron-ion
bremsstrahlung process while propagating downward along the flare loop.
The other part of the stored magnetic energy could be used to drive
coronal mass ejections and heat the plasma \citep{2017ApJ...836...17A}.
However, it is poorly understood exactly how much of this stored energy
goes into different processes. Understanding the location and properties
of the energy release is the key to better understanding the
acceleration mechanism \citep{2011Holman,2017Benz}. These flare-accelerated
electrons also produce heating during the collision with ambient plasma.
One step towards understanding the energetics of this complex
energy-conversion system is to follow the temporal evolution of the
flare-accelerated electrons.

As X-rays are produced in an optically thin environment, they serve as
one of the most direct methods to study the electron acceleration
processes in flares \citep{2011Kontar}. 
Unlike radio or optical, Hard X-Ray (HXR) photons 
are weakly affected by the media during their propagation to the detector 
[with an important exception to albedo
\citep{1978ApJ...219..705B,2006A&A...446.1157K,2020A&A...642A..79J}].
The HXR flux is also proportional to the nonthermal electron density, 
providing us with a comparatively direct way 
to probe the nonthermal electron 
distributions.
This is therefore no surprise that the study of flare-accelerated
electrons using Hard X-Ray observations has been one of the most active
fields of research in solar physics for a long time
\citep{1985Dennis,1988Dennis,2011Holman,2011Kontar}. 
However, for completeness, we note that HXR producing electrons 
are affected by the return current loses along the legs of flaring loop \citep{1977Knight,2005zharkova,2006zharkova,2012karlicky}, Langmuir waves \citep{2009Hannah,2013hannah}, 
electron re-acceleration \citep{2012A&A...539A..43K,2023ApJ...943L..23K} 
and by the presence of turbulent fluctuations along the flaring loop 
\citep{2016Bian,2024ApJ...977..246E} that might complicate the inference 
of the injected/accelerated electron distribution.

The Reuven Ramaty High Resolution Solar Spectroscopic Imager (RHESSI;
\cite{2002SoPh..210....3L}) spacecraft and the Spectrometer/Telescope
for Imaging X-Rays (STIX; \cite{2020krucker}) instrument onboard the
Solar Orbiter \citep[SolO]{2020muller} spacecraft provide us with a
unique opportunity to venture into the physics of solar flares through
their high resolution spectra and spectroscopic imaging. HXR
observations of the coronal sources and chromospheric footpoints
\citep{2002SoPh..210..373B,2002SoPh..210..383A,2003Emslie,2008Kontar,2022A&A...659A..60M}
shows support for the well-known thick-target model
\citep{1971Brown,1972Sv} for flares. In the cold-target model, it is
assumed that the accelerated electrons loss their energy due to the
binary Coulomb collisions with the ambient electrons allowing inference
the injected (accelerated) electron rate spectrum \citep{1971Brown}.
Over the past few decades, this cold-target consideration has been one
of the crucial milestones to understand and model the electron
acceleration mechanism in solar flares \citep{1971Brown,1972Sv,
1976Lin,2011Holman}. Despite a wide-range success of the cold-target
model, there are a few caveats in this approach. Observations show that
during solar flares, the coronal plasma temperature ($T$) often reach over
$10^7$K, suggesting a warm-target condition for the electrons with
energy of a few thermal energies ($k_{\rm B}T$, $k_{\rm B}$ is the Boltzmann constant) \citep[see e.g.][ for discussion]{2003Emslie}.
This therefore necessitates a significant change 
in the description of nonthermal electrons.
The warm-target model developed by \cite{2015Kontar} 
takes care of the electrons with an energy of a few $k_{\rm B}T$.
The HXR spectrum modeled within the warm-target approach takes care of
the contribution from the electrons in the cold chromosphere and the
warm corona. The warm-target model includes the spatial diffusion of the
nonthermal electrons which thermalize inside the target, conserving the
total number of injected electrons \citep{2019Kontar,2024Yingjie}. The
nonthermal electron spectrum is a power-law
\citep{1967SoPh....2..171A,1971ApJ...164..151K,1988Dennis,1992A&A...265..278T,2011Holman}
and the energy contained in these accelerated electrons is strongly
dependent on the low-energy cut-off of the electron distribution,
$E_{\rm c}$ \citep[see][as a review]{2019ApJ...881....1A}.
Underestimating the value of $E_{\rm c}$ even by a factor of two can
cause an order of magnitude ($2^3$) overestimation of the energy content in accelerated electrons of typical electron 
spectrum $\propto E^{-4}$ ($E$ is the electron energy). 
Therefore, $E_{\rm c}$ plays a central role in
flare energetics \citep{2019Kontar,2019ApJ...881....1A}. 
This is crucial in order to quantify the energy 
that is associated with the electron acceleration
process in solar flares
\citep{2012ApJ...759...71E,2019ApJ...881....1A,2020Kumar,2023ApJ...952...75G,2023ApJ...946...53S,2023ApJ...958...18P,2024ApJ...973...96A}.
Under the cold-target framework, $E_{\rm c}$ remains undetermined 
\citep[e.g. 100\% uncertainty at $2\sigma$ level][]{2013Ireland}, 
therefore, the overall energetics of accelerated electrons becomes poorly constrained. 
On the other hand, the warm-target model can determine $E_{\rm c}$ with $\approx 7\%$ uncertainty at the $3\sigma$ level \citep[see Fig 1 of][]{2019Kontar}. This permits an accurate estimation of the energy content in accelerated electrons. Therefore, we use the warm-target model for the current study. 

Moreover, investigating the time evolution of electrons 
within warm-target model is another key aspect, which is currently poorly investigated. For example, \citet{1982machado} shows the importance of the temporal
evolution of continuous energy release in flares. 
\citet{2017Kontar} and \citet{2021ApJ...923...40S} 
demonstrate complex temporal and spatial
evolution of the nonthermal electron distribution and plasma turbulence
believed to be responsible for particle energization.
\cite{2004GrigisBenz} report that the soft-hard-soft behavior of the
spectral index is not only true for the rise-peak-decay phases of a
flare but is even more pronounced in the subpeaks. They also find an
empirical power-law dependence between the spectral index and the
nonthermal photon flux \citep[see also][]{2003Holman}. This analysis was
performed using the cold-target model, which cannot determine the
low-energy cut-off of the nonthermal electron distribution.

In this study, we investigate the time evolution of three well-observed
flares within the framework of the warm-target model using X-ray
observations of the two flares with RHESSI and one with SolO
spacecrafts. This paper aims at addressing the following three
questions: i) How does the low-energy cut-off of accelerated electron
distribution vary as a function of time throughout the flare,
ii) How does the total power of the flare change with time, and iii) How
does the excess thermal emission measure from the nonthermal electrons
contribute to the total emission measure during the time evolution of
the flares.

The paper is organized as follows: \S~\ref{S:warm-target} briefly
describes the warm-target model in association with the nonthermal
parameters mentioned above. 
The application of the warm-target model to each of the three flares
along with the results are described in \S~\ref{S: flares application}.
Finally, \S~\ref{S: summary} contains the summary.

\section{Accelerated electrons in a warm-target model}
\label{S:warm-target}

In this section, we briefly outline the warm-target model. The density
averaged mean electron flux spectrum under this model can be written as
\citep{2015Kontar,2019Kontar}
 \begin{equation}\label{eq: nVF}
     \langle n V F \rangle (E) \approx \Delta EM \sqrt{\frac{8}{\pi m_e}}
	 \left( \frac{E}{(k_{\rm B}T)^{3/2}}\right) e^{-E/k_{\rm B}T} + \frac{E}{K} \int_{E}^{\infty} \dot{N}(E_0)dE_0
 \end{equation}
where, $E$ is the electron energy, $n$ is the electron number density
(cm$^{-3}$), $m_{\rm e}$ is the mass of an electron in g, $K = 2\pi e^4
\ln \Lambda \simeq 2.6 \times 10^{-18}$ (keV$^2$~cm$^2$)
\citep{1978Emslie} is a constant used to describe the electron Coulomb
collisions with $e$ (esu) as the charge of electrons and $\ln
\Lambda\simeq 20$ as the Coulomb logarithm, $k_{\rm B}$ is the Boltzmann
constant, $T$ is the plasma temperature in Kelvin, and $\Delta EM$ is
the excess emission measure coming from the thermalization of nonthermal
electrons in a flare.

The first term in the right-hand side (RHS) of Equation (\ref{eq: nVF})
represents the contribution from the thermalized electrons, while the
second term depicts the contribution from the electrons without
thermalization. For $E\gg k_{\rm B}T$, the first term in the RHS goes to
zero, and Equation (\ref{eq: nVF}) therefore reduces to the cold-target
form. Equation (\ref{eq: nVF}) with an assumption that $\dot{N}(E_0)$
follows a power law at high energies, is incorporated in the  Object
SPectral EXecutive (OSPEX) software package
\citep{2020ascl.soft07017S,2020ascl.soft07018T} as the function
`$f_{thick-warm}$'. The detailed use of such routines are demonstrated
by
\citet{2019Kontar,2021ApJ...908..111X,2023ApJ...952...75G,2024ApJ...964..145J,2024Yingjie}.
The mean electron flux for a given electron energy $E$ (Equation
(\ref{eq: nVF})) convolved with the electron-ion bremsstrahlung
cross-section, $Q(\epsilon, E)$, where $\epsilon$ and $E$ are the photon
and electron energy, respectively, predicts the X-Ray photon flux
spectrum at a distance $d$ from the flare:
\begin{equation}
    I(\epsilon) = \frac{1}{4\pi {d^2}} \int_{\epsilon}^{\infty} \langle n V F \rangle (E) Q(\epsilon, E) dE.
\end{equation}
To determine the nonthermal electron flux spectrum, we assume the
injected spectrum (the rate spectrum of accelerated electrons) as,
\begin{equation}
\label{eq: N}
    \dot{N}(E) = \dot{N_0} \frac{\delta - 1}{E_{\rm c}} \left( \frac{E}{E_{\rm c}} \right)^{-\delta}\,,
\end{equation}
where, $\dot{N_0}$ is the total injected electron rate (electrons
s$^{-1}$), $\delta$ is the spectral index of the electron distribution,
and $E_{\rm c}$ (keV) is the low-energy cut-off of the nonthermal
electron spectrum. Integrating over the nonthermal range of energy
($E_{\rm c}$ to $\infty$), we get the total rate of the injected
electrons in the system,
\begin{equation}
\label{eq: N0}
    \dot{N_0} = \int_{E_c}^{\infty} \dot{N}(E) dE.
\end{equation}
The total power, $P$ (keV s$^{-1}$) associated with the nonthermal
electrons becomes
\begin{equation}
\label{eq: P}
    P = \int_{E_{\rm c}}^{\infty} E \dot{N}(E) dE = \frac{\delta -1}{\delta -2} \dot{N_0}E_c.
\end{equation}
For a given power law electron spectrum, $\dot{N}(E)$, $P$ is crucially
dependent on the estimation of $E_{\rm c}$. The injected electrons
become partially thermalized in the warm plasma of flare coronal loop.
These electrons therefore produce a quantifiable thermal emission. This
additional or excess thermalized emission measure from the nonthermal
electrons is \citep{2019Kontar}:
\begin{equation}
 \label{eq: delta_EM}
     \Delta EM \approx \frac{\pi}{K} \sqrt{\frac{m_e}{8}} (k_{\rm B}T)^2 \frac{\dot{N_0}}{(E_{\rm min})^{1/2}},
 \end{equation}
where $E_{\rm min} \approx 3k_{\rm B}T(5\lambda / L)^4$ and $\lambda = (k_{\rm B}T)^2 /
(2Kn)$ is the electron collisional mean-free path. $L$ is the half-loop
length of the flare. $\lambda$ can be determined from the warm plasma
properties in the solar corona and by the gradual escape of electrons
into the cold chromosphere \citep{2015Kontar,2019Kontar}. $\Delta EM$
characterizes the nonthermal electron distribution that thermalizes in
the hot coronal part of the flare loop and escapes into the
chromosphere. In order to investigate the contribution of $\Delta EM$
with respect to the total emission measure ($EM = EM_0 + \Delta EM$), we
define the ratio $R$ as
\begin{equation}
 \label{eq: R}
     R = \frac{\Delta EM}{(\Delta EM + EM_0)},
 \end{equation}
where $EM_0$ is the thermal emission measure of the warm plasma in the
flare loop before electron injection. The injected electrons with energy
$E \leq \sqrt{2KnL}$ thermalize before reaching the chromosphere,
therefore increasing the density (and $\Delta EM$) of the warm plasma in
the flaring loop. $R$, therefore suggests the accumulation of these
extra electrons that are getting thermalized from the injected ones.
$\Delta EM \propto \dot{N_0}$, implying the higher injection rate of
electrons leads to larger $R$, e.g. injected electrons that did not
diffusively escape into the chromosphere.

The warm-target model uses a combination of X-ray spectroscopy and
imaging to quantify both the thermal and nonthermal parameters. Imaging
is required to determine the plasma temperature ($T$) of the coronal
source, the half-loop length ($L$) of the flare, and the number density
($n$) of the plasma.

\begin{figure}
	
\includegraphics[width=0.55\textwidth, height = 0.42\textwidth]{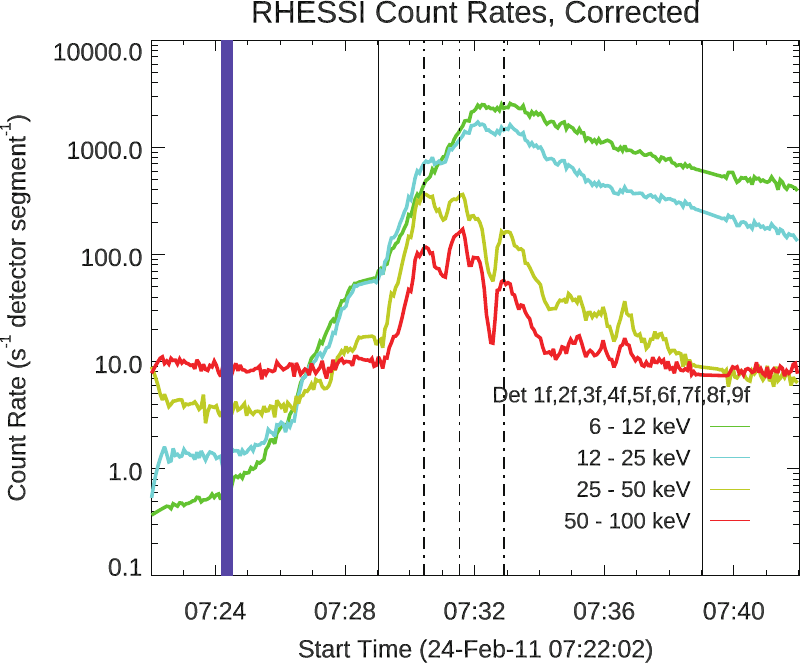}
\includegraphics[width=0.44\textwidth]{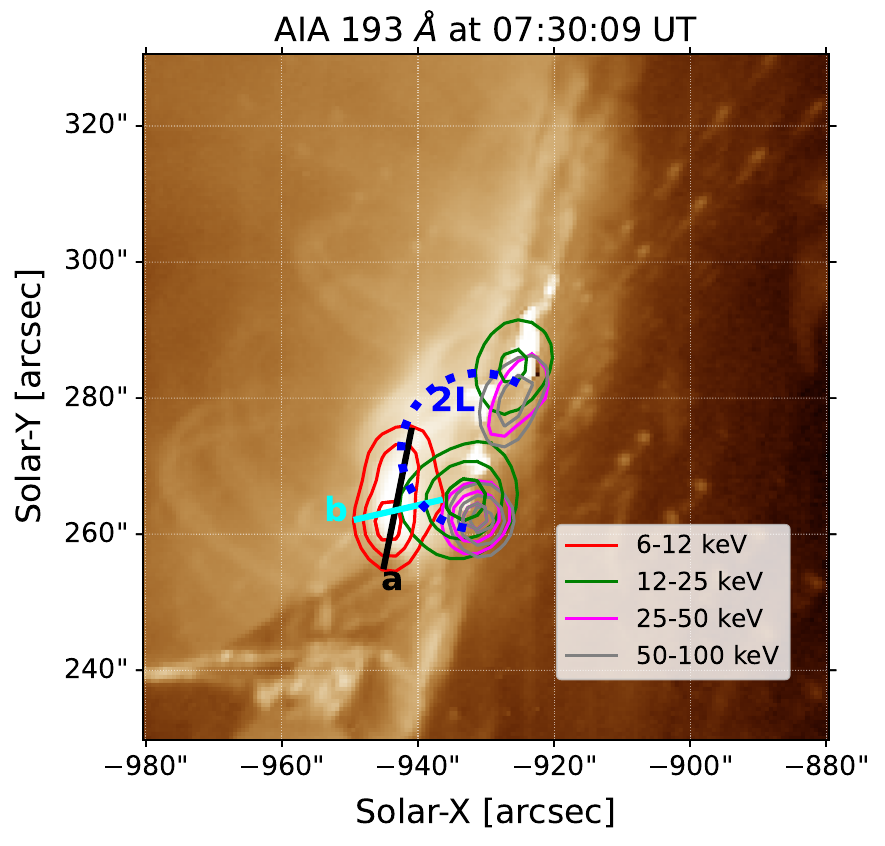}
	
	\caption{The left panel shows the RHESSI light curve of the solar
	flare February 24, 2011. Four energy windows (6-12, 12-25, 25-50,
	and 50-100 keV) of our interest are shown in this light curve. The
	region between the two solid black vertical line represent the time
	window of our study for this flare. The dark blue region before
	the left black solid vertical line shows the chosen background. The three black dotted vertical lines represent the three HXR peaks of this flare. The right
	panel shows the AIA 193~{\AA} image overlapped with RHESSI X-ray CLEAN
	contour maps (50\%, 70\%, and 90\% of the maximum). The coronal source region and two
	foot-points of this flare are clearly visible in this image. The
	blue dotted arc of length $2L$ represent the full loop length.
	Therefore, $L$ represents the half-loop length of this flare. The dimensions
	of the coronal source region with 50\% contour, marked $a$ and
	$b$, are used to estimate the area of the contour. This area is  used to determine the volume (V) of the coronal source.}
\label{Fig: Fl20110224_ltc}
\end{figure}

\begin{figure}
\includegraphics[height = 0.42\textwidth , width=0.5\textwidth]{flare_20110224_lightcurve_10_c.pdf}
{\tikz\node[coordinate](start1){};}
\includegraphics[width=0.49\textwidth]{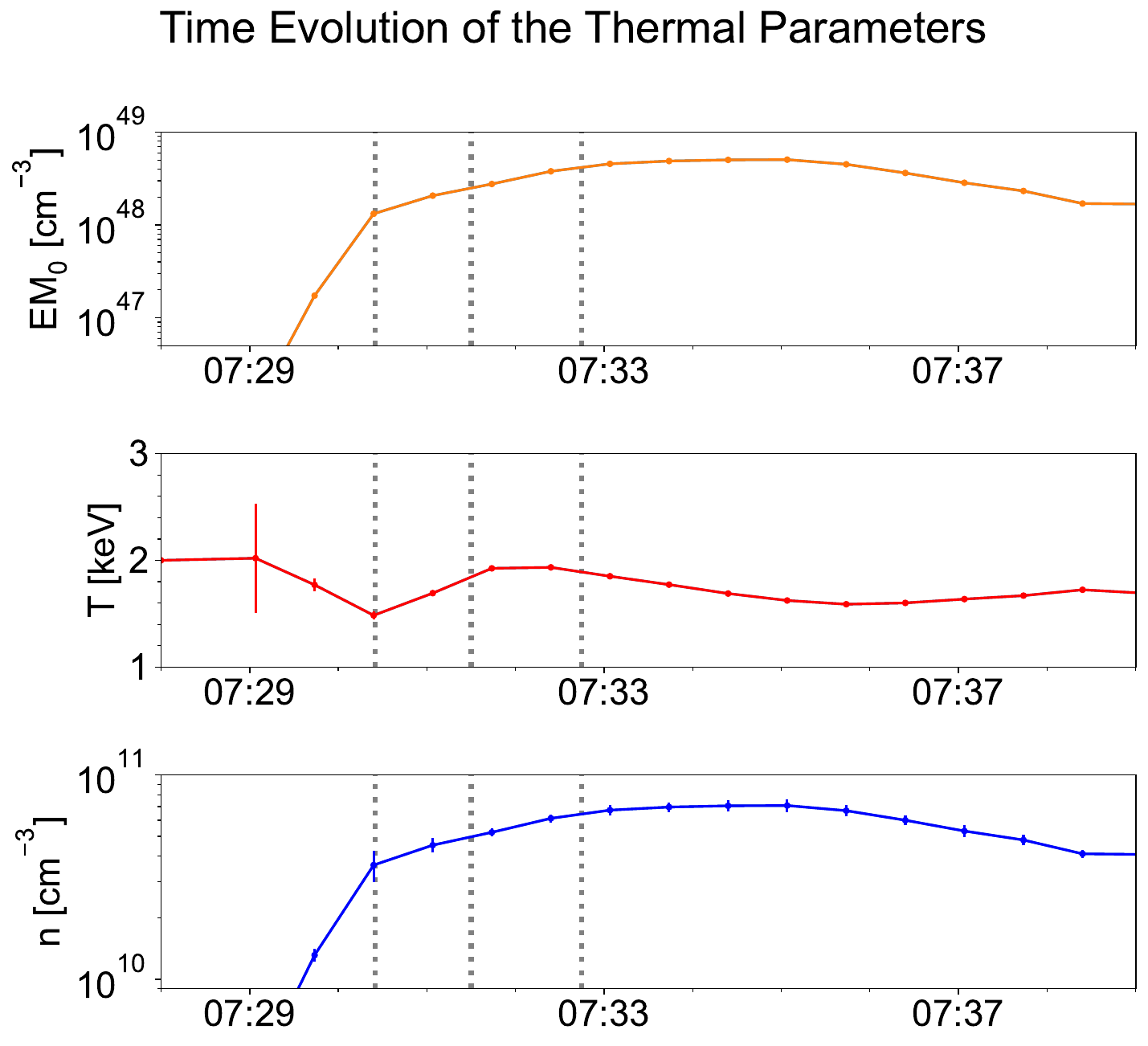}
\includegraphics[width=0.5\textwidth]{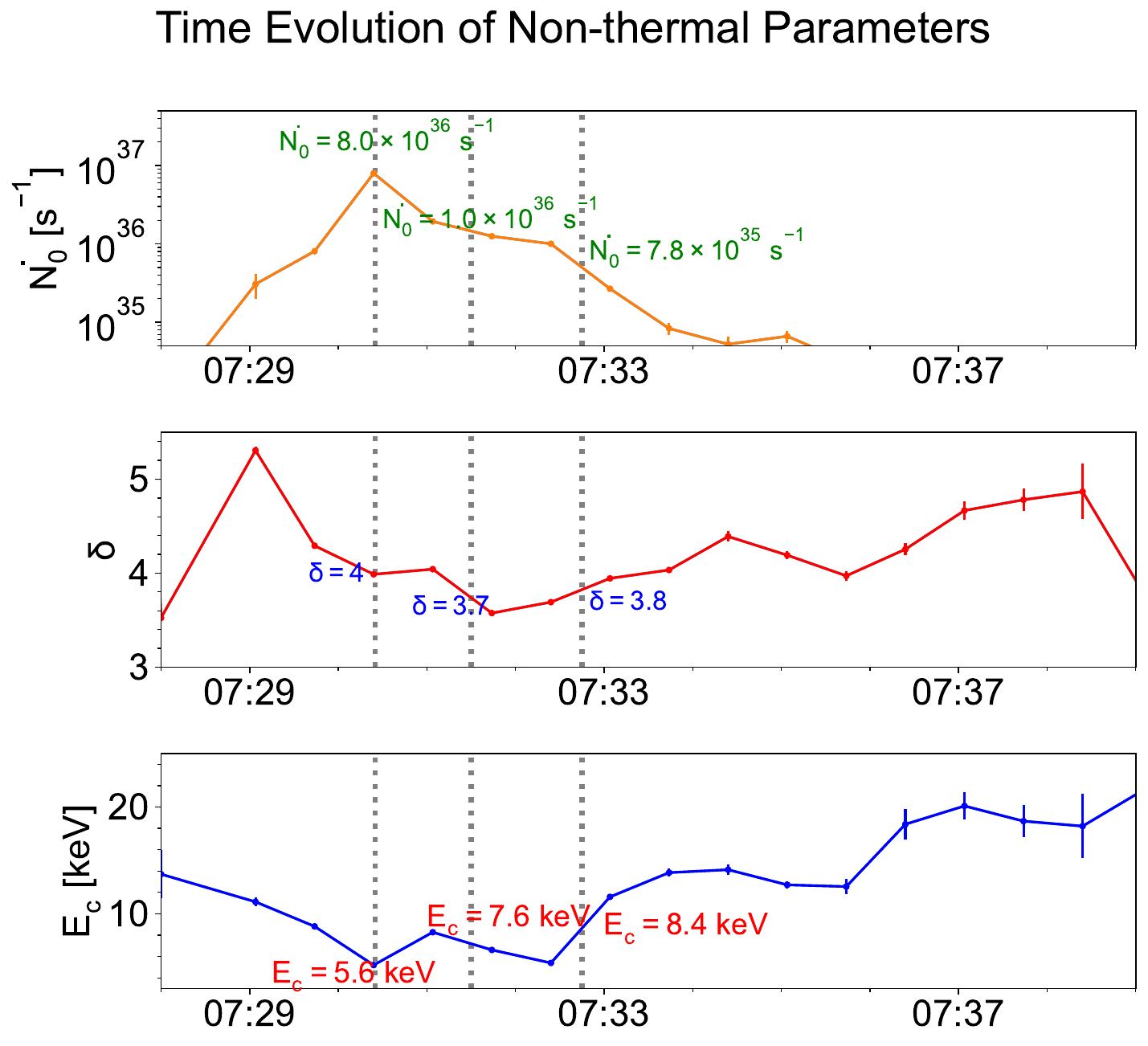}
{\tikz\node[coordinate](end1){};}
\includegraphics[width=0.49\textwidth]{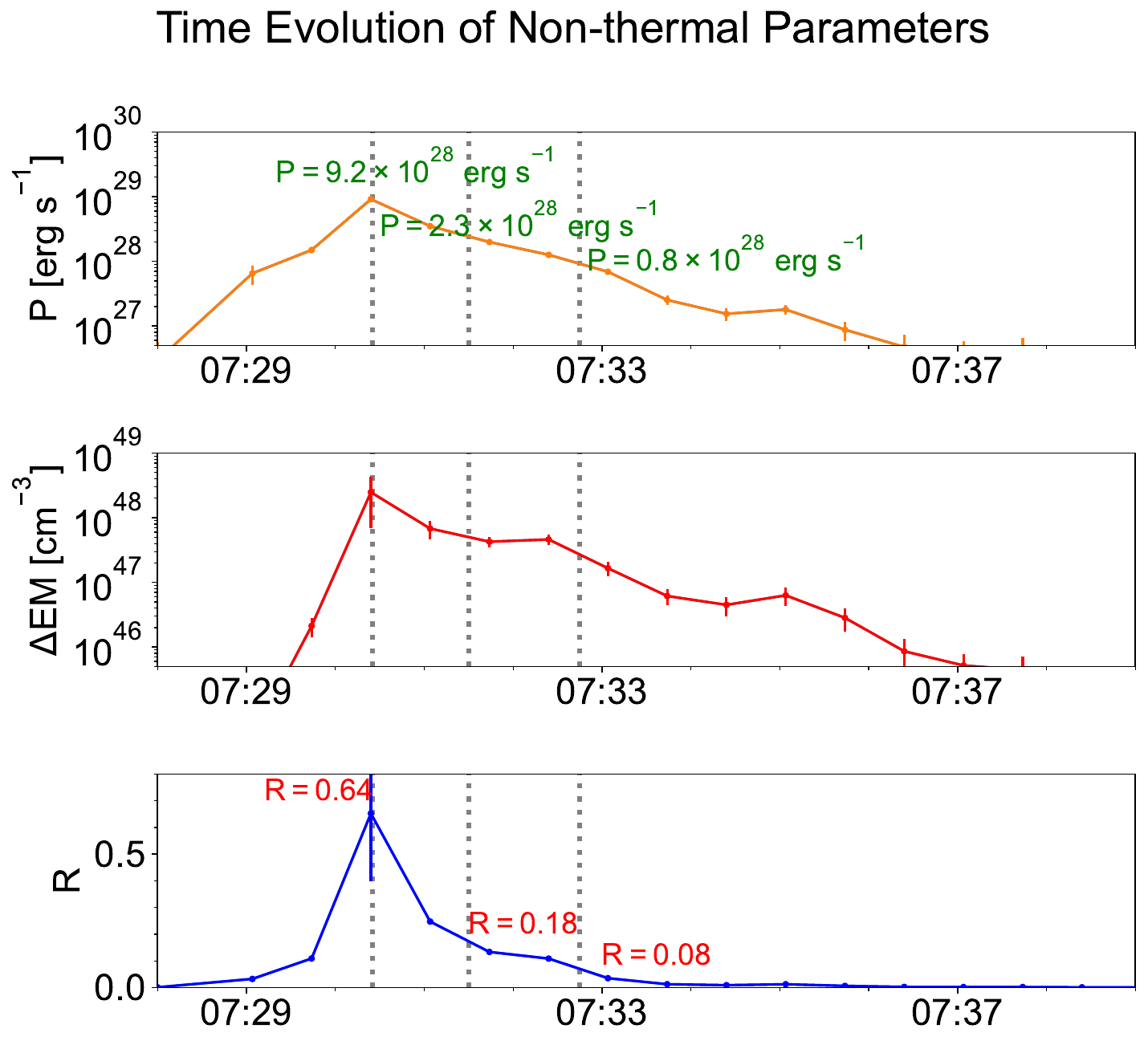}

	\caption{The top left panel shows the light curve of the RHESSI
	flare February 24, 2011 as a reference. The region between two
	solid black vertical lines represents the full window where we perform
	the time evolution for this flare. The top right panel shows the
	temporal evolution of thermal parameters derived by fitting the
	$f_{\rm vth}$ function in the pre-burst X-ray spectrum. The
	thermal parameters include: thermal emission measure ($EM_0$,
	cm$^{-3}$), plasma temperature ($T$, keV), and plasma number density
	($n$, cm$^{-3}$). The bottom two panels display the time evolution
	of nonthermal parameters determined using warm-target model to the burst 
	X-ray spectrum. These include: the total rate of injected electrons
	($\dot{N_0}$, s$^{-1}$), power law index ($\delta$), low energy
	cut-off ($E_c$, keV), total power of nonthermal electrons ($P$, erg
	s$^{-1}$), excess thermal emission measure from nonthermal electrons
	($\Delta EM$, cm$^{-1}$), and the ratio between $\Delta EM$ and
	total emission measure ($EM_0 + \Delta EM$), $R$.}
\begin{tikzpicture}[overlay, remember picture, -latex, color=blue!55!red, yshift=150ex, shorten >=0pt, shorten <=0pt, line width=0.03cm]
  \path[-] ($(start1)+(-7.8cm,-0.86cm)$) edge [out=35, in=215]
  ($(end1)+(-4.87cm,9.22cm)$);
\end{tikzpicture}
\begin{tikzpicture}[overlay, remember picture, -latex, color=blue!55!red, yshift=150ex, shorten >=0pt, shorten <=0pt, line width=0.03cm]
  \path[-] ($(start1)+(-0.2cm,-0.86cm)$) edge [out=124, in=296]
  ($(end1)+(-1.2cm,9.22cm)$);
\end{tikzpicture}
    
\label{Fig: Flare_Feb24}
\end{figure}

\section{Time evolution of solar flare accelerated electrons}
\label{S: flares application}

This section aims at the application of the warm-target model to three
well-observed solar flares observed by the RHESSI and SolO (STIX
instrument) spacecrafts. 
The first two flares
(February 24, 2011, and May 15, 2013) are observed by the RHESSI, while
the third flare (March 28, 2022) is observed by the SolO. The GOES M
class flares are sufficiently strong to provide countrate for reliable
imaging and spectroscopy and are free from the pile-up effects (high
detector lifetime ($> 85\%$), \citet{2002SoPh..210...33S}). On the
other hand, the May 15, 2013 is a GOES X1.2 class flare and have
associated pile-up corrections. Of these three flares, two (February 24,
2011 and March 28, 2022) are limb flares and therefore free from the
Compton back scattering of the flare hard X-rays, also known as the albedo
correction
\citep{2006A&A...446.1157K,2013SoPh..284..405D,2019SoPh..294..105A,2024ApJ...964..145J},
which changes the observed spectra for on-disk flares. The third flare
(the May 15, 2013) is not a limb flare, and the albedo corrections
\citep{2006A&A...446.1157K} are included in the fit. These three flares are selected because they are close to the limb events \citep{2011Battaglia,2017Kontar, 2023purkhart,2024Yingjie} and show clear signatures of two X-ray foot-points and loop-top coronal source throughout their temporal evolution. We investigate the
temporal evolution of various parameters associated with thermal and
nonthermal electrons throughout the flare using a combination of X-ray
imaging and spectroscopy using the OSPEX software package
\citep{2020ascl.soft07017S} in the SSW IDL \citep{1998sswidl} suite for
the spectral fitting of the flares.

The thermal properties corresponding to each flare are determined from
the isothermal fit (using the $f_{\rm vth}$ function) to the X-ray
spectrum and HXR imaging. The warm-target model assumes that the thermal
properties ($EM_0$ and $T$) are estimated using the isothermal ($f_{\rm
vth}$) fit to the spectrum in the pre-burst \citep{2019Kontar,2024Yingjie}.

\subsection{Flare: February 24, 2011}
\label{S: Feb24}
The flare has three major HXR peaks at 07:30:25 UT, 07:31:30 UT, and
07:32:45 UT (marked by three gray dotted vertical lines in the light
curve shown in the left panel of Figure~\ref{Fig: Fl20110224_ltc}),
respectively, and soft X-ray (SXR) peaks around 07:35:00 UT. This flare
occurred on the eastern limb of the Sun. During the observation of this
flare, the spacecraft was in fully operational mode, providing us with the
high quality imaging and spectra.

\begin{figure}
	\centering
	\includegraphics[width= 0.54\textwidth]{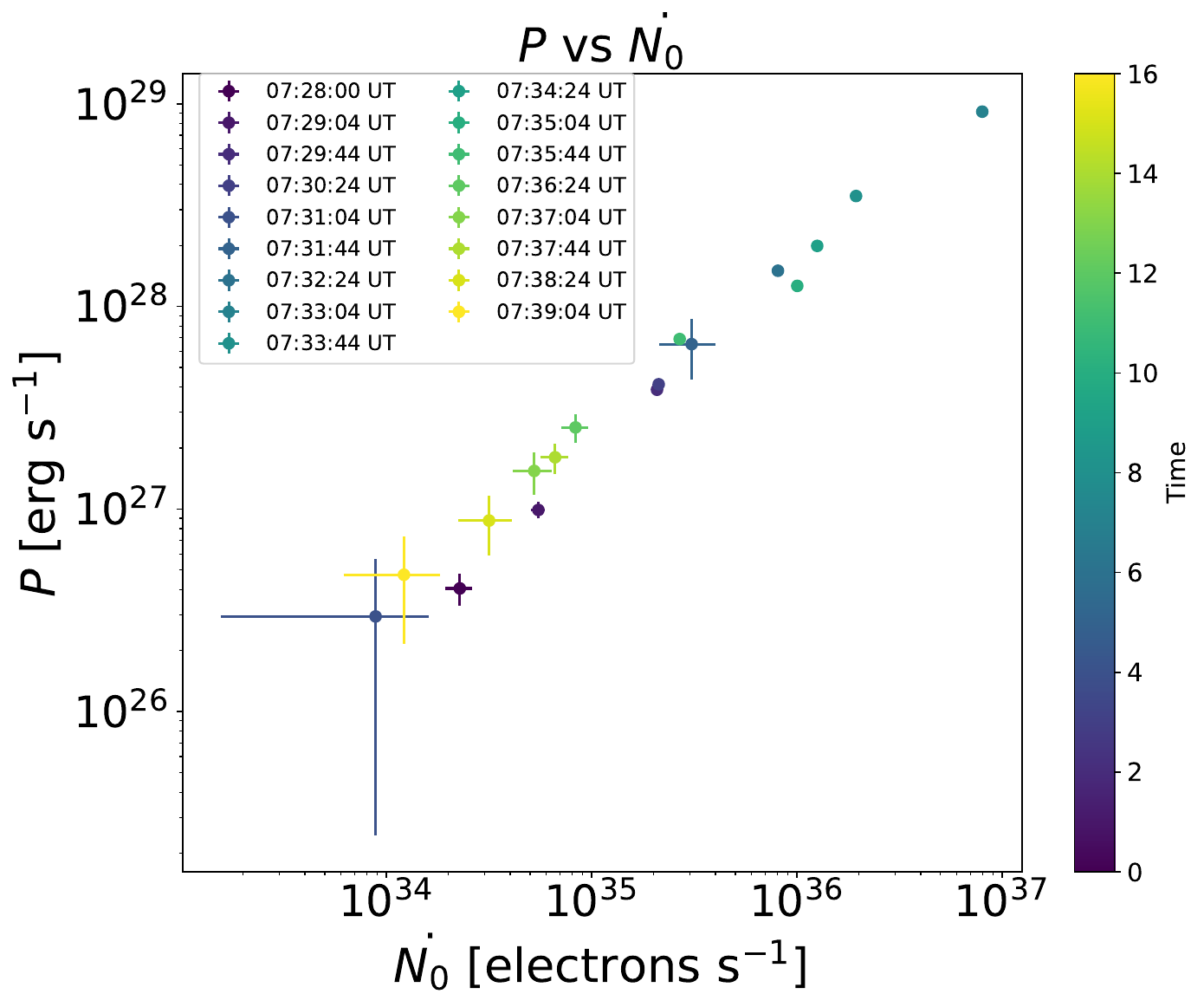}
	\caption{The power of nonthermal electrons ($P$) vs the total rate
	of nonthermal electrons ($\dot{N_0}$) from the Hard X-ray fits for
	the flare February 24, 2011. The legend describes the chosen intervals for this flare.}
\label{Fig: Fl20110224_linear}
\end{figure}

\begin{figure}
	\centering
	\includegraphics[width= 0.55\textwidth]{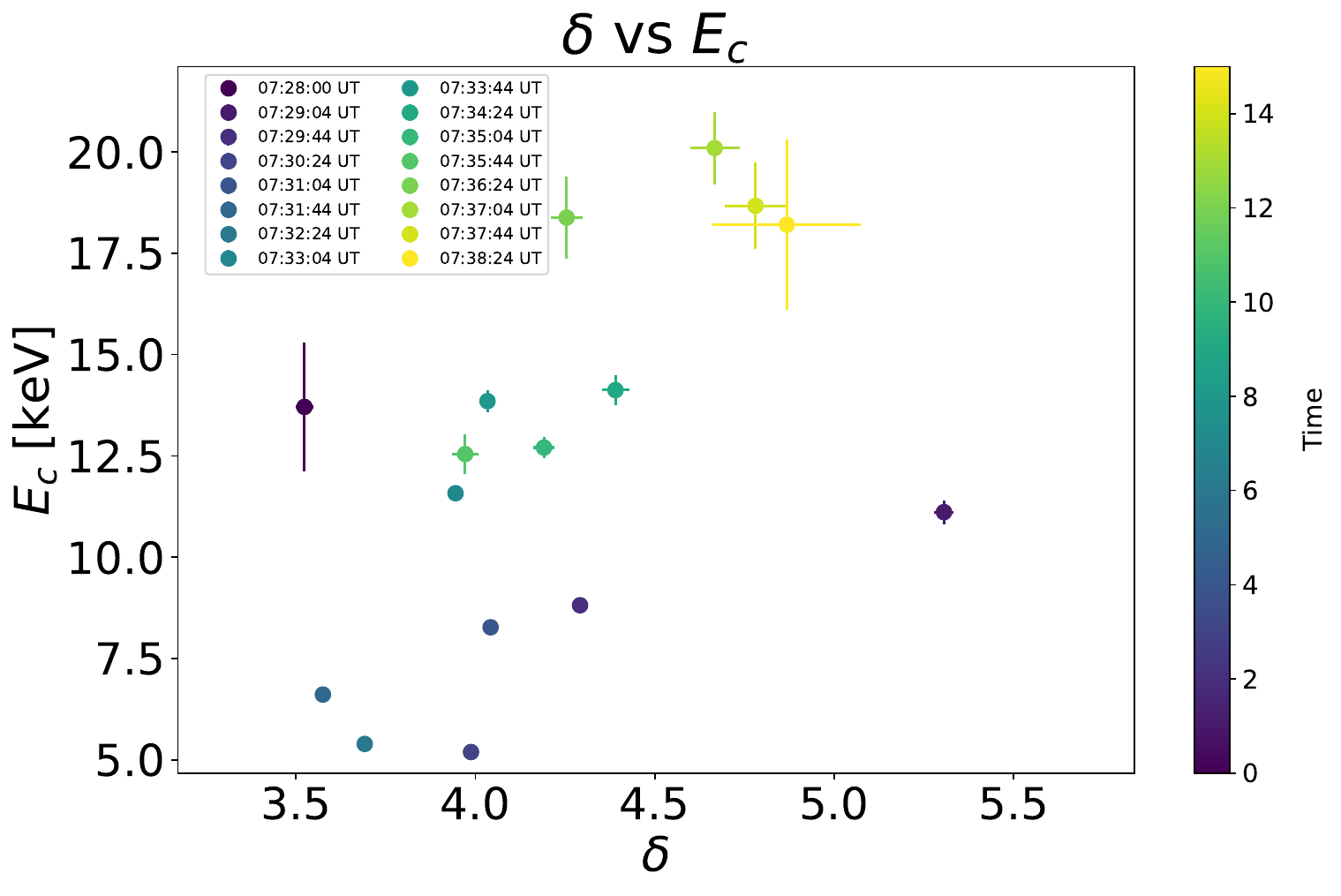}

	\caption{The low-energy cut off of nonthermal electron distribution ($E_{\rm c}$) vs the spectral index ($\delta$) from the Hard X-ray fits for
	the flare February 24, 2011. The legend describes the chosen intervals for this flare. The Pearson's correlation coefficient ($r$) between these two parameters is 0.53 with p-value 0.02.}
\label{Fig: Fl20110224_corr}
\end{figure}

The left panel of Figure~\ref{Fig: Fl20110224_ltc} shows the RHESSI
light curve for the energy windows: 6-12, 12-25, 25-50, and 50-100~keV.
The interval between 07:28:00 UT and 07:39:04 UT (the two black vertical
solid lines shown in the light curve) is chosen to for the temporal
evolution of particle acceleration study. The dark blue window around 07:24
UT represents the selected background for our study. The right panel of
Figure~\ref{Fig: Fl20110224_ltc} shows the 193 {\AA} Atmospheric Imaging
Assembly (AIA, \cite{2012lemen}) image, which overlaps with the 50\%,
70\%, and 90\% of the maximum RHESSI CLEAN \citep{2002hurford} X-ray
contours around the first HXR peak of this flare. The coronal source
region and the two foot-points are clearly visible in this image. In
order to determine the flare half-loop length ($L$), a semicircular arc
that connects the two foot-points of the flare and passes through its
coronal source region is assumed (see the right panel of Figure~\ref{Fig:
Fl20110224_ltc}) We estimate $L = 19 \arcsec \approx 13.9$ Mm using the
50\% of the maximum RHESSI X-ray contour level to determine the coronal
source volume ($V$) of this flare. 

Following the procedure by \citet{2024Yingjie}, the cross-sectional area
of the region in the right panel of Figure~\ref{Fig: Flare_Feb24} is
estimated as $A = \pi (b/2)^2 = 6.2 \times 10^{17}$ cm$^2$ using $a = 20
\arcsec \approx$ 14.5 Mm and $b = 12 \arcsec \approx 8.7$ Mm. The
coronal source volume is, therefore, becomes $V \approx A \times a = 8.8
\times 10^{26}$~cm$^{3}$. Our estimated values of $L$ and $V$ are in
agreement with the findings of \cite{2024Yingjie}. In this study, $L$
and $V$ are assumed to be constant in time. The whole time window
(between two solid black vertical lines in the left panel of
Figure~\ref{Fig: Fl20110224_ltc}) is split into equal time intervals, each
of 40 seconds (hereafter abbreviated as `s'), avoiding the attenuators. The selection of the time window is based on the average
diffusion time of thermal electrons (using Eq. 11 of \cite{2019Kontar}).
The warm-target model requires the estimation of the thermal properties
($EM_0$, $T$) of the warm-plasma using a pre-burst \citep{2019Kontar,2024Yingjie}. Each 40~s interval serves as the
pre-burst for the 40 s interval immediately after it to determine the
thermal properties for the next interval. The time profiles of $EM_0$
and $T$ are determined by fitting the X-ray spectrum corresponding to
each pre-burst with the $f_{\rm vth} + f_{\rm thick2}$ using the energy
range 6-100 keV. Then, the time profile of thermal electron number
density ($n$) is estimated using $n = \sqrt{EM_0 / V}$. The time
profiles of $EM_0$, $T$, and $n$ are shown in the top right panel of
Figure~\ref{Fig: Flare_Feb24}. The three dotted vertical lines represent the three
major HXR peaks of the flare.

Once the thermal parameters are estimated, $f_{\rm thick-warm}$
(warm-target model) with $f_{\rm vth}$ used at the energy range 6-100
keV to each 40~s interval. The average thermal electron diffusion time
for this flare is $\approx 60$ s (using Equation~11 of
\cite{2019Kontar}), so the thermal parameters vary slowly within this
time interval. While fitting, the magnitudes of $T$ and $EM_0$ are kept
fixed in $f_{\rm vth}$ function according to the values estimated in the
pre-burst spectra. The elemental abundance are assumed to be coronal
(default value of 1 in OSPEX
\citep{2012ApJ...748...52P,2013Landi,2022ApJ...930...77S}) for all three
flares analyzed in this study. The bottom-left panel of Figure~\ref{Fig:
Flare_Feb24} shows the temporal evolution of the nonthermal electron
distribution parameters $\dot{N_0}$, $\delta$, and $E_{\rm c}$ deduced.

We find that total rate of injected electrons ($\dot{N_0}$) has its
maximum at the first HXR peak (07:30:25 UT) with $\dot{N_0} = (8.0 \pm
0.2) \times 10^{36}$ s$^{-1}$. The magnitudes of $\dot{N_0}$ are $1.0
\times 10^{36}$ s$^{-1}$ and $7.8 \times 10^{35}$ s$^{-1}$ at the second
and third HXR peaks, respectively. We note that we do not have data
points at the second and third HXR peaks, these magnitudes are estimated
using the linear interpolation method. Therefore, we do not have error
bars at these HXR peaks.


The time evolution of power-law index ($\delta$) shows the usual
soft-hard-soft profile around the peaks
\citep[e.g.][]{1977benz,1988kosugi}. The magnitudes of the low-energy
cutoff ($E_{\rm c}$) at these three peaks are $5.6 \pm 0.1$, $7.6$, and
$8.4$ keV, respectively. These $E_{\rm c}$ values are lower than the
regions immediately before and after the HXR peaks (e.g., at 07:29:00
UT, $E_{\rm c} = 11.1 \pm 0.4$ keV and at 07:33:00 UT, $E_{\rm c} = 11.6 \pm 0.2$ keV, see
the bottom-left panel of Figure~\ref{Fig: Flare_Feb24}), suggesting a
drop of magnitude of $E_{\rm c}$ around the HXR peaks. Therefore, the
time profile of $E_{\rm c}$ shows a high-low-high trend around the HXR
peaks during the time evolution of the flare. On the other hand,
$\dot{N_0}$ shows a low-high-low trend around the HXR peaks. We note
that both $E_{\rm c}$ and $\delta$ have a drop in their magnitudes
around the HXR peaks, suggesting that the hardness of the accelerated
electron spectra is related to the lower value of $E_{\rm c}$ and
therefore a higher rate in injected electrons.
We estimate the Pearson's correlation coefficient ($r$) between the time profiles of $E_{\rm c}$ and $\delta$ to be 0.53 with p-value 0.02 (see Fig~\ref{Fig: Fl20110224_corr}). A p-value $\leq 0.05$ suggests a 95\% confidence in the estimated value of $r$ \citep{pvalue}. \\
The power $P$ shows a linear relationship with the time profile of
$\dot{N_0}$ on the log-log scale (see Figure~\ref{Fig:
Fl20110224_linear}). The maximum magnitude of $P$ ($= (9.2 \pm 0.2)
\times 10^{28}$ erg s$^{-1}$) coincides with the maximum of $\dot{N_0}$
($= (8.0 \pm 0.2) \times 10^{36}$ electrons s$^{-1}$), implying that the
total power of the nonthermal electrons peaks at the first HXR peak of
this flare. After the first HXR peak, $P$ gradually decays as the flare
progresses. The middle and the bottom plots of the middle-right panel of
Figure~\ref{Fig: Flare_Feb24} describe the time profiles of $\Delta EM$
(Equation (\ref{eq: delta_EM})) and $R$ (Equation (\ref{eq: R})). At the
first HXR peak, both $\Delta EM$ and $R$ have their maximum with values
$2.3 \pm 0.8 \times 10^{48}$ cm$^{-3}$ and $0.64 \pm 0.20$,
respectively. $R = 0.18$, and 0.08 at the second (07:31:30 UT) and third
(07:32:45 UT) HXR peaks, respectively, showing a decrease of $R$ after
the first HXR peak. This suggests that $\Delta EM$ has a $64 \pm 20 \%$
contribution to the total emission measure ($EM$) at the first HXR peak
of this flare. This contribution drops rapidly as the flare progresses
in time, leaving a 18\% contribution at the second peak and only a 8\%
contribution at the third peak. We note that this contribution is $>
10\%$ only between the times 07:29:45 UT and 07:32:30 UT. $R$ can be
represented as the fraction of thermalized injected electrons in a
flare. 

\begin{figure}
\includegraphics[width=0.52\textwidth]{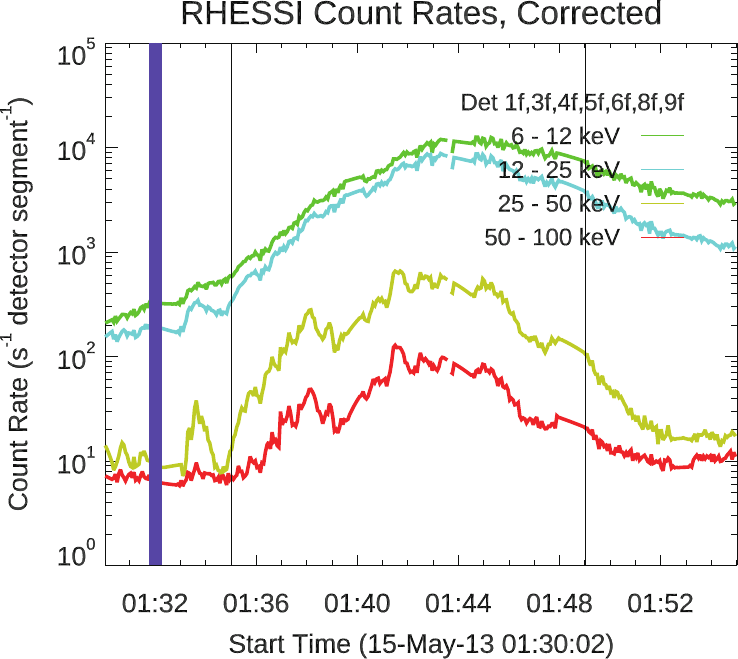}
\includegraphics[width=0.47\textwidth]{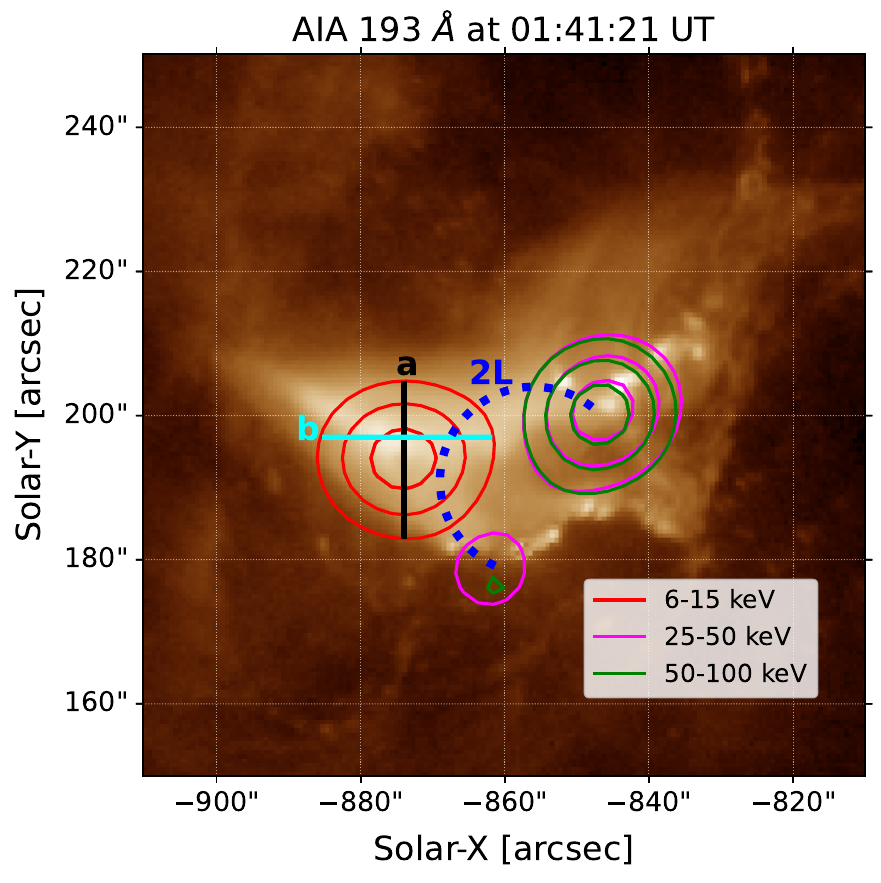}	
	\caption{The left panel shows the RHESSI light curve of the solar flare May 15, 2013. Four energy windows (6-12, 12-25, 25-50, and 50-100 keV) of our interest are shown in this light curve. The region between the two black solid vertical lines represent the time window of our study for this flare. The dark blue region before the left black vertical line shows the chosen background. The right panel shows the AIA 193 {\AA} image overlapped with RHESSI CLEAN X-ray contour maps (50\%, 70\%, and 90\% of the maximum). The coronal source region and two foot-points of this flare are clearly visible in this image. The blue dotted arc of length $2L$ represent the full loop length. Therefore, $L$ is the half-loop length of this flare. The dimensions of the coronal source region with 50\% contour, marked $a$ and $b$, are used to estimate the area of the contour. This area is used to determine the volume (V) of the coronal source.}
\label{Fig: Fl20130515_ltc}
\end{figure}

\begin{figure}
\includegraphics[width=0.5\textwidth, height=0.38\textwidth]{flare_20130515_lightcurve_7_c.pdf}
{\tikz\node[coordinate](start1){};}
\includegraphics[width=0.49\textwidth]{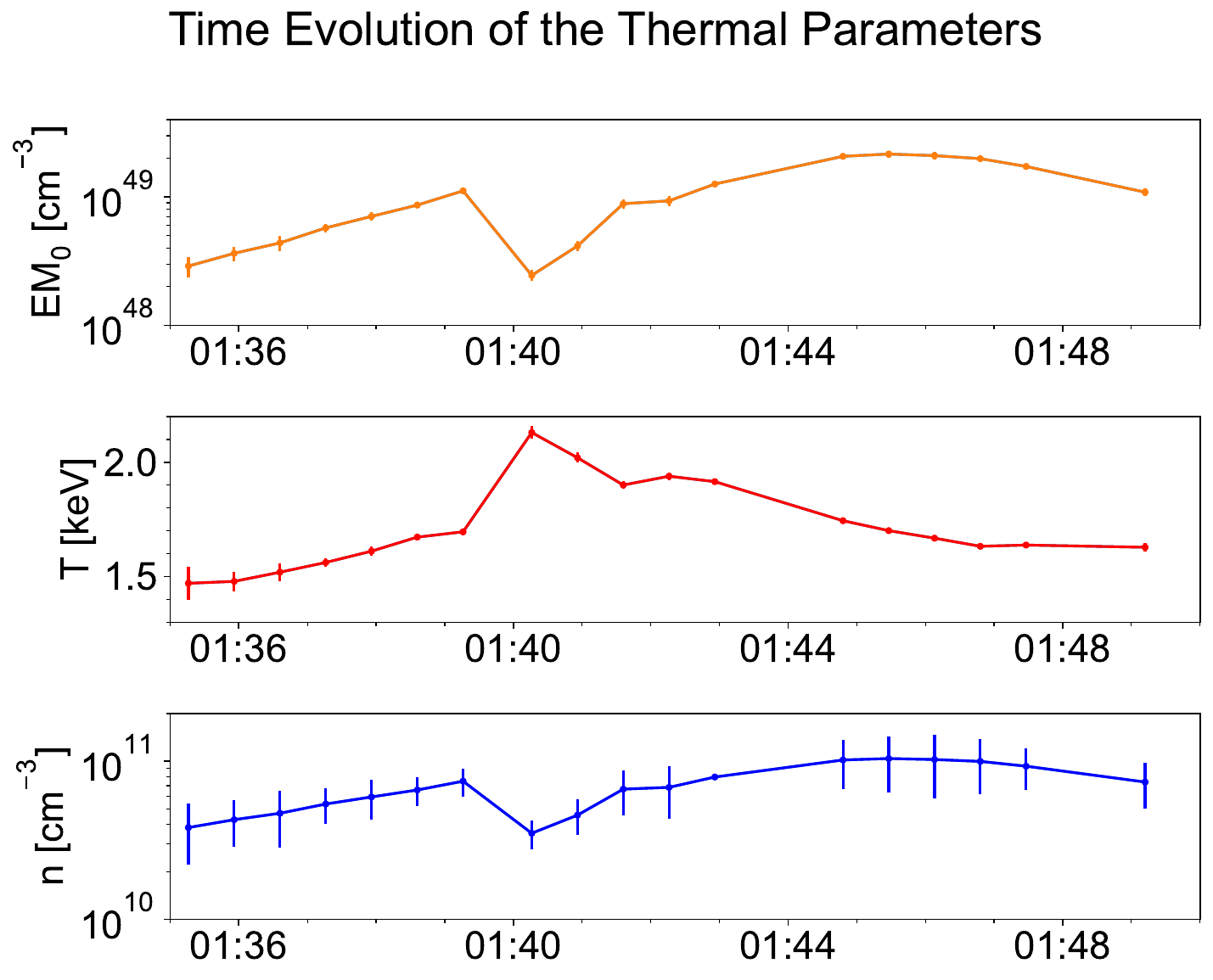}
\includegraphics[width=0.5\textwidth]{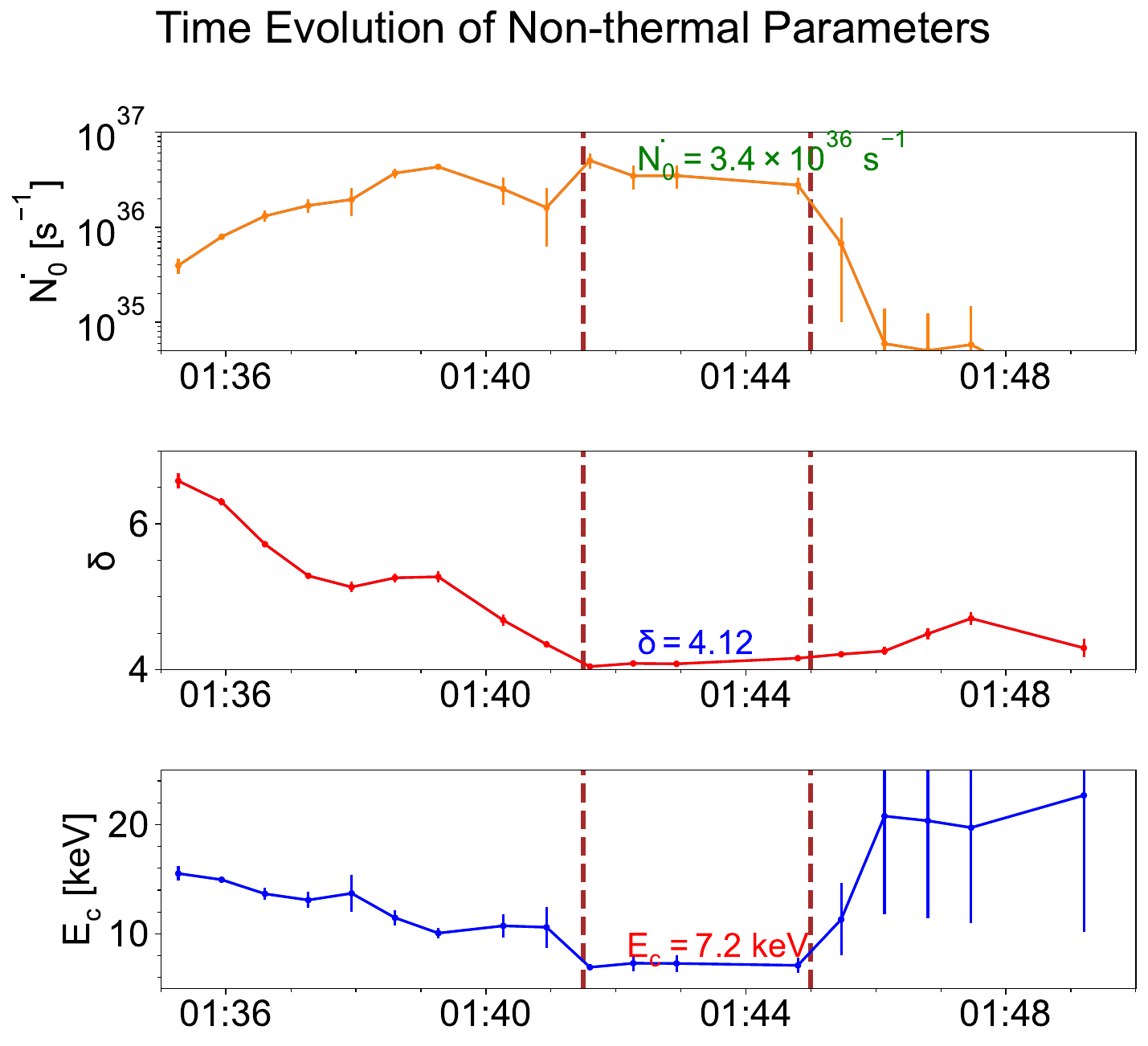}
{\tikz\node[coordinate](end1){};}
\includegraphics[width=0.49\textwidth]{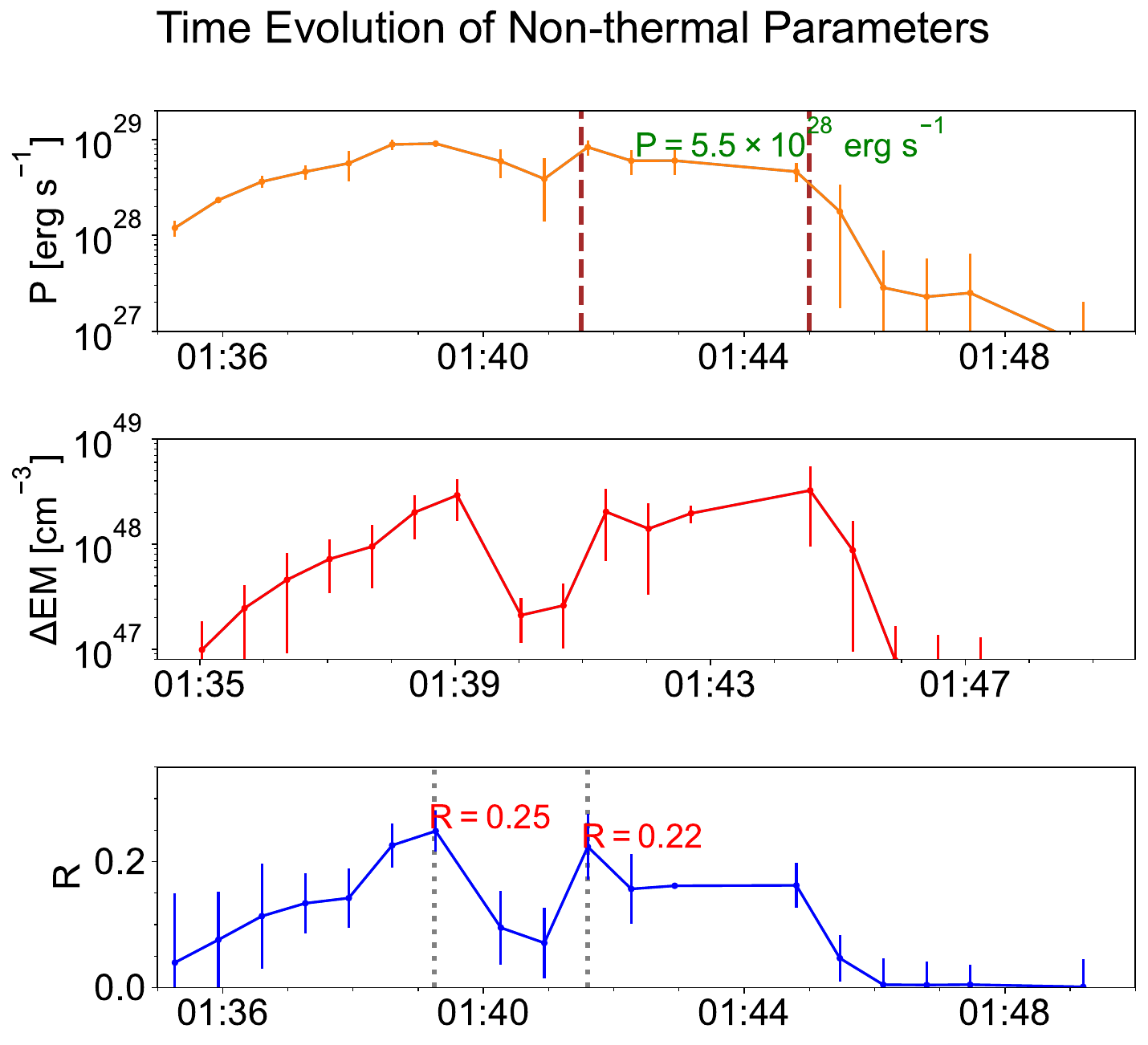}
\caption{The top left panel shows the light curve of the RHESSI flare
	May 15, 2013 as a reference. The region between two black solid vertical
	lines represents the full window where we perform the time evolution
	for this flare. The top right panel shows the temporal evolution of
	thermal parameters derived by fitting the $f_{\rm vth}$ function in
	the pre-burst X-ray spectrum. The thermal parameters include:
	thermal emission measure ($EM_0$, cm$^{-3}$), plasma temperature
	($T$, keV), and plasma number density ($n$, cm$^{-3}$). The bottom
	two panels display the time evolution of nonthermal parameters
	determined using warm-target model to the X-ray spectrum of the burst. These
	include: the total rate of injected electrons ($\dot{N_0}$,
	s$^{-1}$), power law index ($\delta$), low-energy cut-off ($E_c$,
	keV), total power of nonthermal electrons ($P$, erg s$^{-1}$),
	excess thermal emission measure from nonthermal electrons ($\Delta
	EM$, cm$^{-3}$), and the ratio between $\Delta EM$ and total
	emission measure ($EM_0 + \Delta EM$), $R$. We note an approximate constancy of $\dot{N_0}$, $\delta$, $E_{\rm c}$, and $P$ in the temporal region between two brown dashed vertical lines. The two grey dotted vertical lines show two peaks in $R$. }

\begin{tikzpicture}[overlay, remember picture, -latex, color=blue!55!red, yshift=150ex, shorten >=0pt, shorten <=0pt, line width=0.03cm]
  \path[-] ($(start1)+(-7.8cm,-1.05cm)$) edge [out=46, in=238]
  ($(end1)+(-6.3cm,9.2cm)$);
\end{tikzpicture}
\begin{tikzpicture}[overlay, remember picture, -latex, color=blue!55!red, yshift=150ex, shorten >=0pt, shorten <=0pt, line width=0.03cm]
  \path[-] ($(start1)+(-0.22cm,-1.05cm)$) edge [out=128, in=310]
  ($(end1)+(-2.0cm,9.2cm)$);
\end{tikzpicture}

\label{Fig: Flare_May15}
\end{figure}

\subsection{Flare: May 15, 2013} \label{S: May15}

The second RHESSI flare in our event list is the GOES X1.2 class solar
flare that took place on May 15, 2013 in the NOAA solar active region
11748. The left panel of Figure~\ref{Fig: Fl20130515_ltc} shows the
RHESSI light curve (corrected counts) for the energy windows: 6-12,
12-25, 25-50, and 50-100 keV. This flare has a single HXR peak unlike
the other two flares (February 24, 2011 and March 28, 2022) discussed in
this study. The region between the two black vertical solid lines in the
RHESSI light curve (see left panel of Figure~\ref{Fig: Fl20130515_ltc}),
which implies the time between 01:35:16 UT and 01:49:00 UT, mark the
time window of study for this flare. The dark blue region in front of the
flare displays the chosen background for the analysis. The right panel
of Figure~\ref{Fig: Fl20130515_ltc} shows the 193 {\AA} AIA image
over-plotted with the 50\%, 70\%, and 90\% of the maximum CLEAN RHESSI
X-ray contours at 01:41:21 UT. The contours at energy range 6-15 keV
(red) mark the coronal loop-top source, while the contours at energy
ranges 25-50 keV (magenta) and 50-100 keV (green) show the two
footpoints of this flare. We draw a semicircular arc (see the blue
dotted arc marked ``2L" in the right panel of Figure~\ref{Fig:
Fl20130515_ltc}) to determine the half-loop length ($L$) of this flare.
This arc joins the two footpoints of the flare and passes through its
coronal source region. We find $L = 22 \arcsec \approx 16$ Mm, which is
in agreement with the value estimated by \cite{2017Kontar}. Like the
flare previously analyzed flare in \S~\ref{S: Feb24} (February 24,
2011), we use the 50\% of the maximum RHESSI X-ray contour to estimate
the coronal source volume ($V$) for this flare. We estimate the
dimensions of this contour as $a = 20 \arcsec \approx 14.5$ Mm (see
Figure~\ref{Fig: Fl20130515_ltc}), and $b = 23 \arcsec \approx 17$~Mm.
The cross-sectional area of this contour is determined as $A = \pi
(b/2)^2 = 2.1 \times 10^{18}$ cm$^{2}$. The estimated source volume is
therefore, $V \approx A \times a = 3.1 \times 10^{27}$ cm$^3$. We note
that $L$ and $V$ are assumed to be constant throughout the temporal
evolution of this flare. The entire time window of our interest (between
01:35:16 UT and 01:49:00 UT, marked by the region between two black
vertical lines in the light curve of this flare) is split into equal
intervals of 40 s, avoiding attenuators. This selection is based on the average diffusion time
of the thermal electrons throughout the full-time window, which is
estimated to be $\approx 55$~s. Each $40$~s interval is used as the
pre-burst interval for its immediately next 40 s interval (burst
interval) and these pre-burst intervals are used to compute the plasma
thermal properties for its associated burst interval. We fit the X-ray
spectra for these pre-burst intervals with $f_{\rm vth} + f_{\rm
thick2}$ under the energy range 6-100 keV. The fits provide us with the
time profile of $EM_0$ and $T$ which are displayed in the top and middle
plots of the top-right panel of Figure~\ref{Fig: Flare_May15}. We determine
the time profile of $n$ using $n = \sqrt{EM_0 / V}$, which is shown in
the third plot of the top-right panel of Figure~\ref{Fig: Flare_May15}. The
time evolutions of $T$ and $n$ are in agreement with the findings of
\cite{2017Kontar}.

\begin{figure}
	\centering
	\includegraphics[width=0.54\textwidth]{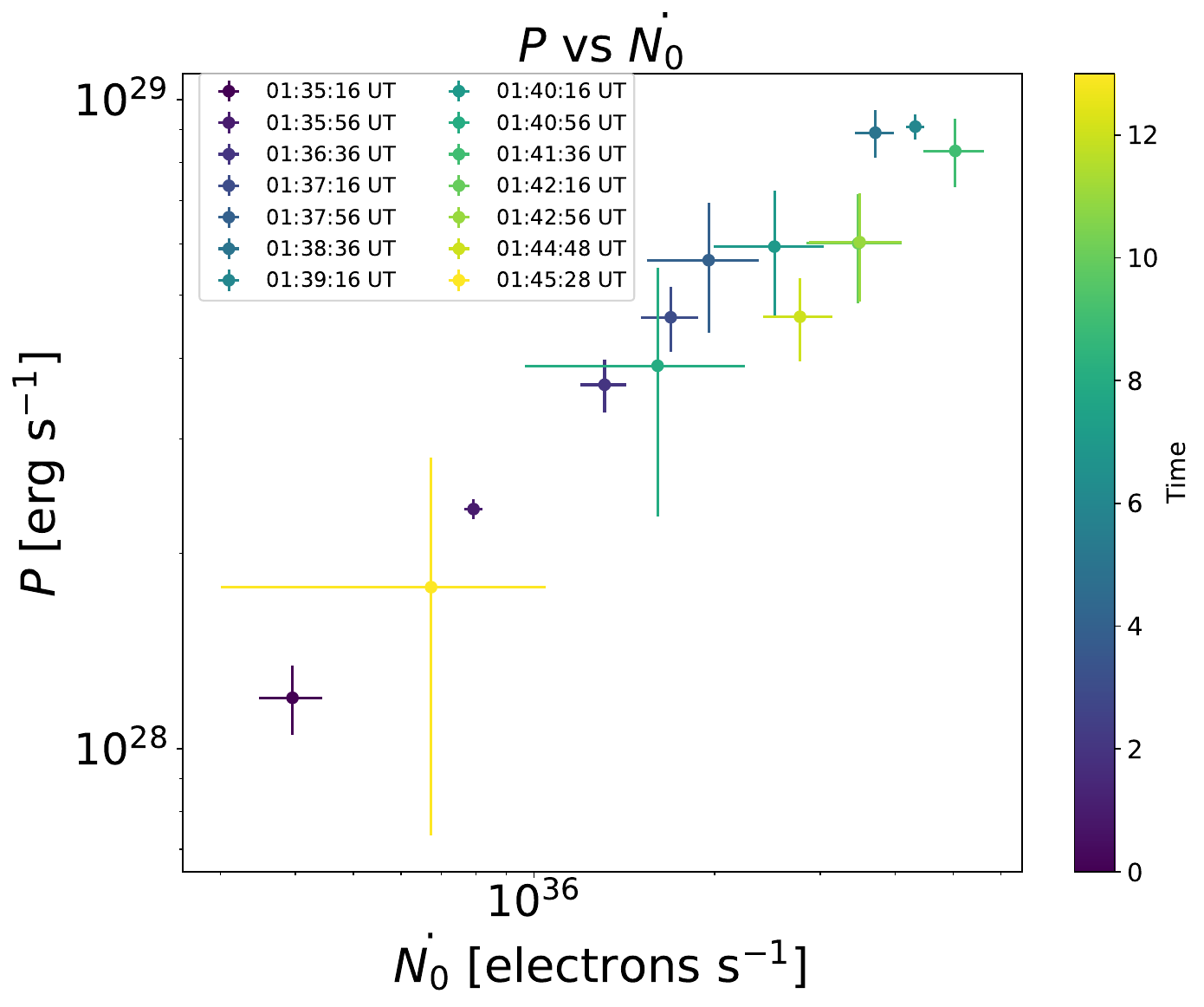}
	\caption{The power of nonthermal electrons ($P$) vs the total rate
	of nonthermal electrons ($\dot{N_0}$) from the Hard X-ray fits for
	the flare May 15, 2013. The legend shows the selected intervals between 01:35:16 UT and 01:45:28 UT for this flare. The last few intervals (01:46:08 UT to 01:49:00 UT) are not included in this plot due to large error bars in $\dot{N_0}$ and $P$ (see Figure~\ref{Fig: Flare_May15}).}
\label{Fig: F2_linear}
\end{figure}

\begin{figure}
	\centering
	\includegraphics[width=0.55\textwidth]{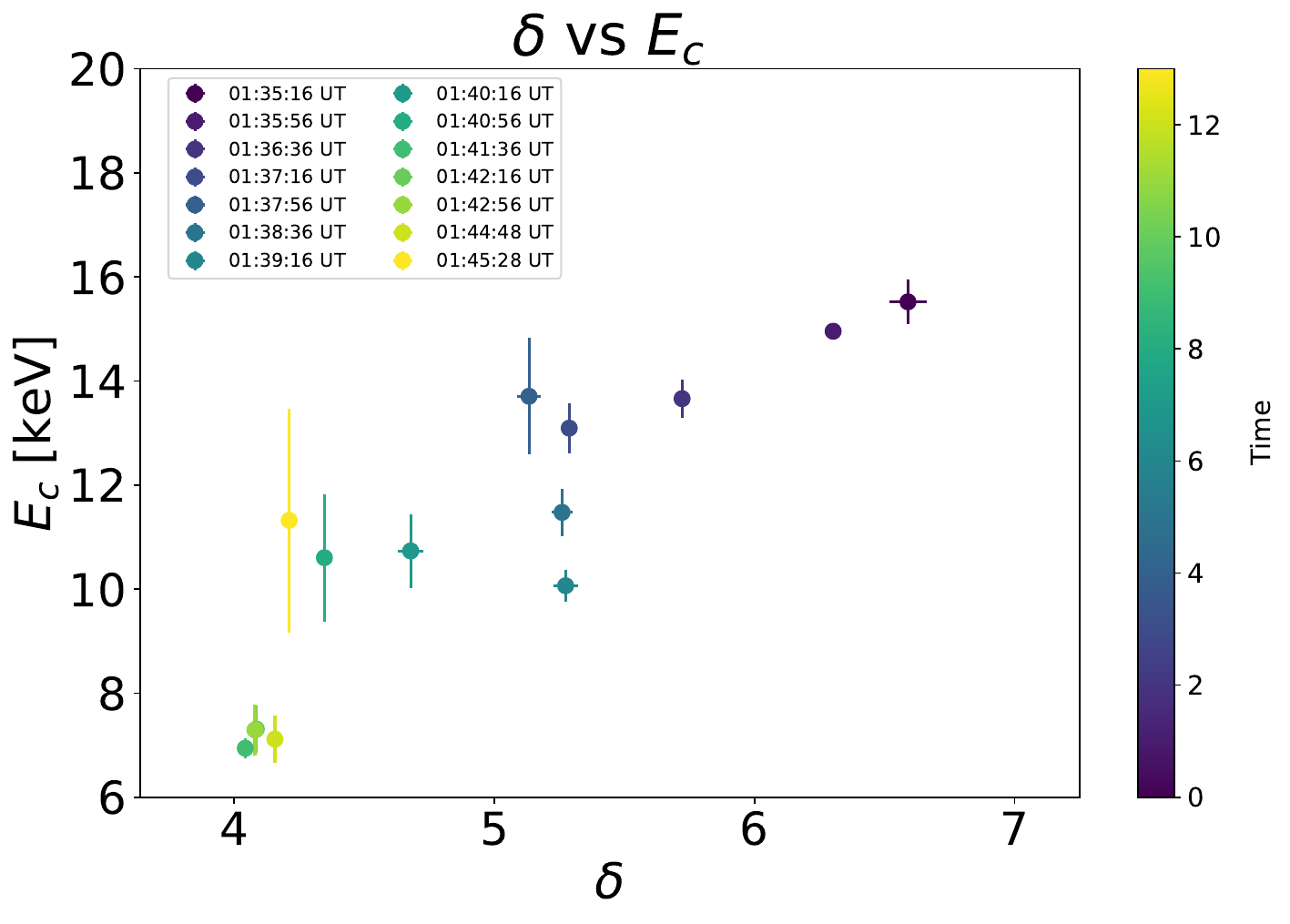}

	\caption{ The low-energy cut off of nonthermal electron distribution ($E_{\rm c}$) vs the spectral index ($\delta$) from the Hard X-ray fits for
	the flare May 15, 2015. The legend describes the chosen intervals for this flare. The Pearson's correlation coefficient ($r$) between these two parameters is 0.39 with p-value 0.02. The legend shows the selected intervals between 01:35:16 UT and 01:45:28 UT for this flare. The last few intervals (01:46:08 UT to 01:49:00 UT) are not included in this plot due to large error bars in $\dot{N_0}$ and $P$ (see Figure~\ref{Fig: Flare_May15}).}
\label{Fig: F2_corr}
\end{figure}

 After the estimation of the thermal parameters, as already explained in
\S~\ref{S: Feb24} for the flare February 24, 2011, we estimate the
nonthermal parameters by fitting $f_{\rm vth} + f_{\rm thick-warm}$ to
the X-ray spectra associated with the burst intervals in the energy range
6-100 keV. The pile-up and isotropic albedo corrections are included in this
analysis. The time profiles of $\dot{N_0}$, $\delta$, and $E_{\rm c}$
are shown in the three panels of the bottom-left panel of Figure~\ref{Fig:
Flare_May15}. We note that between 01:41:30 UT and 01:45:00 UT (the
region between the brown dotted vertical lines in the bottom-left panel
of Figure~\ref{Fig: Flare_May15}), the magnitudes of $\dot{N_0}$, $\delta$,
and $E_{\rm c}$ remain approximately constant under the associated error
bars. In this time interval, their average values are $(3.4 \pm 0.3)
\times 10^{36}$ s$^{-1}$, $4.12 \pm 0.02$, and $7.2 \pm 0.3$ keV,
respectively. We note that the magnitude of $P$ also remains
approximately constant (the temporal region between two brown dashed
lines in the bottom-right panel of Figure~\ref{Fig: Flare_May15}) under the
error bars with an average value of $(5.5 \pm 0.6) \times 10^{28}$ erg
s$^{-1}$. 
The time profile of
$\Delta EM$ (Equation (\ref{eq: delta_EM})) and $R$ (Equation~\ref{eq:
R}) are described in the middle and bottom plots of the bottom-right
panel of Figure~\ref{Fig: Flare_May15}. They also show an approximate
constancy under their uncertainties within that interval. This apparent
constancy of all these nonthermal parameters between 01:41:30 UT and
01:45:00 UT is likely due to the fact that this flare does not have HXR
impulses/peaks for our studied range of energy (6-100 keV) in this time
window. We also note that the time profile of $R$ has two peaks at
01:39:15 UT and 01:41:36 UT with magnitudes $0.25 \pm 0.09$, and $0.22
\pm 0.11$, respectively, suggesting a $25 \pm 9 \%$ and $22 \pm 11 \%$
contribution of $\Delta EM$ to the total emission measure at these two
times, respectively. The time profile of $\delta$ shows the typical
soft-hard-soft behavior during the impulsive phase of this flare.
Similarly to the flare previously analyzed, we note a high-low-high trend
in the temporal evolution of $E_{\rm c}$. This trend is consistent with the
soft-hard-soft (high-low-high) magnitude of $\delta$. We note that $r$ between the temporal evolutions of $\delta$ and $E_{\rm c}$ is 0.39 with p-value 0.02 (see Fig~\ref{Fig: F2_corr}). The minimum of
$E_{\rm c}$ reflects the minimum of $\delta$ and the maximum of
$\dot{N_0}$, suggesting that the spectral hardness of the nonthermal
electron distribution coincides with the lowest value of low cut-off
energy and the highest value in the rate of injected electrons. The time
evolution of $P$ shows a linear relationship with that of $\dot{N_0}$ on the logarithmic scale (see Figure~\ref{Fig: F2_linear}).

\begin{figure}	
\includegraphics[width=0.55\textwidth]{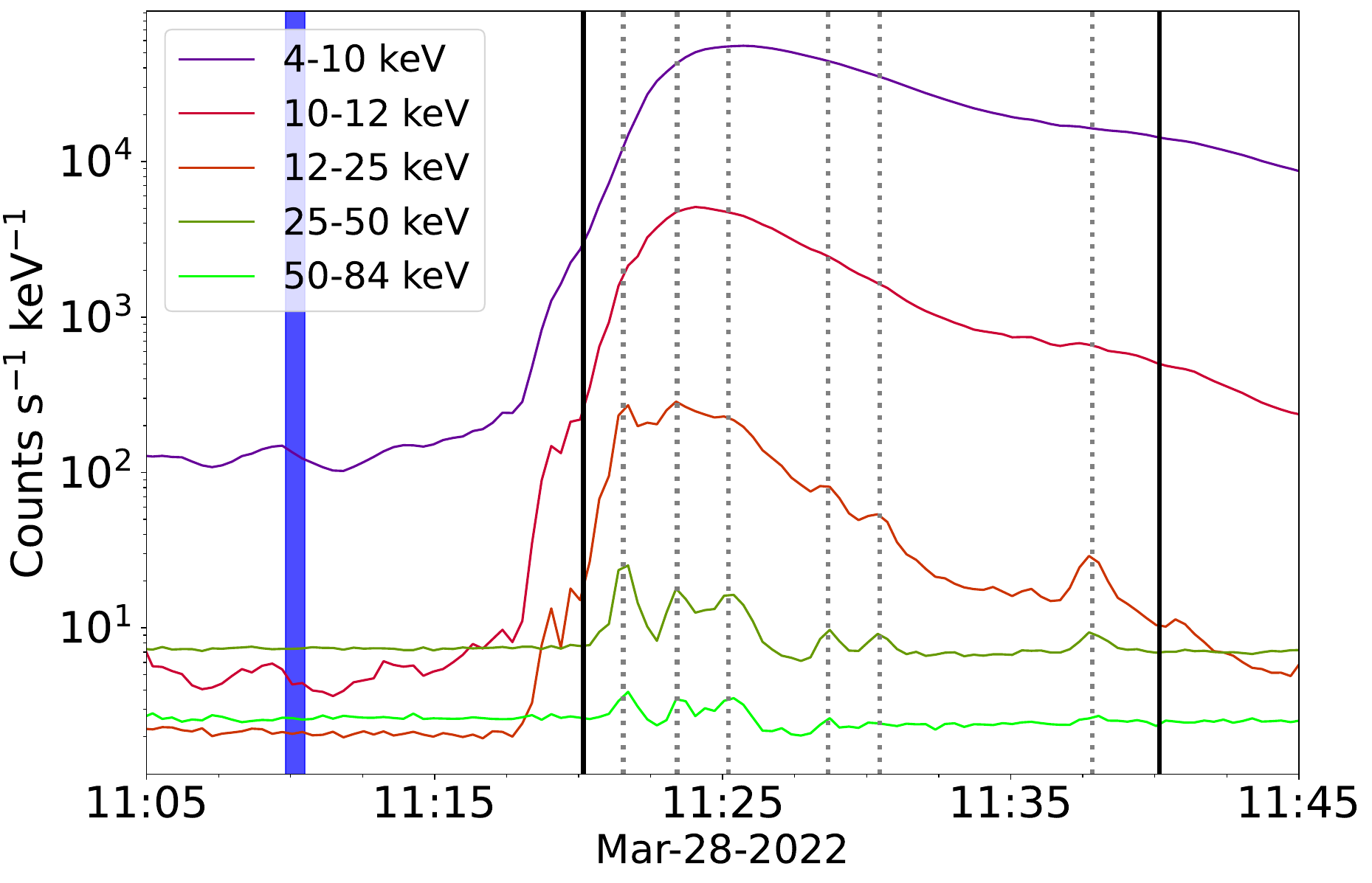}
\includegraphics[width=0.44\textwidth]{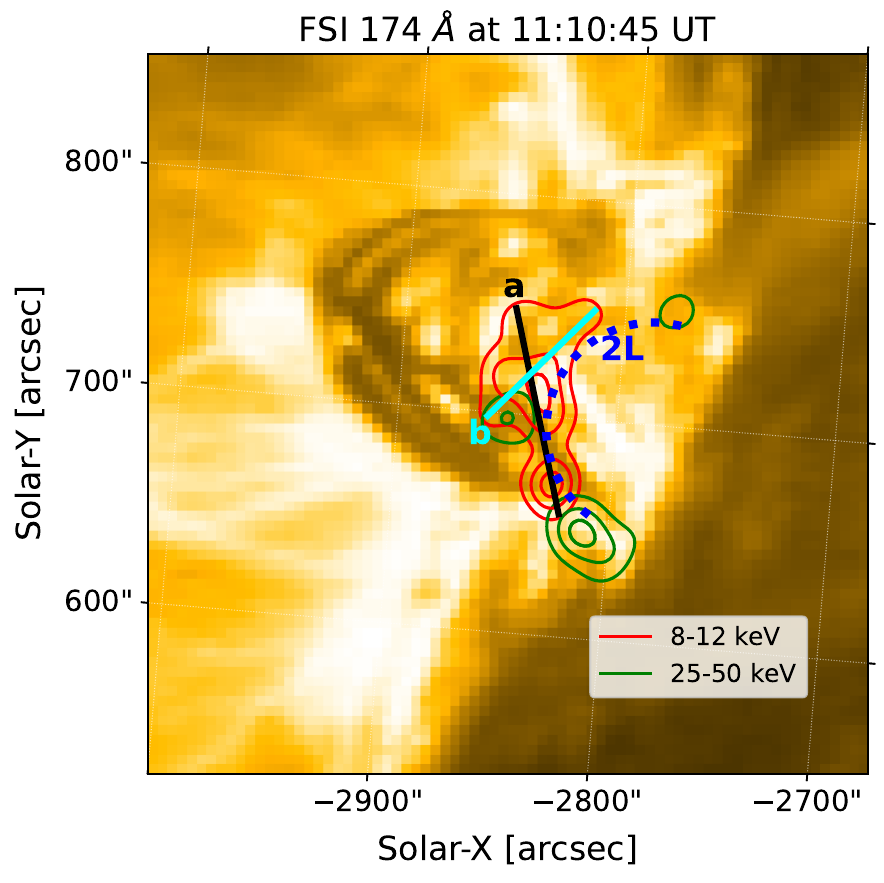}
	\caption{The left panel shows the STIX light curve of the solar
	flare March 28, 2022. Four energy windows (6-12, 12-25, 25-50, and
	50-100 keV) of our interest are shown in this light curve. The
	region between two black solid vertical lines represent the time window
	of our study for this flare. The blue region before the left
	black line shows the chosen background. The six grey dotted vertical lines represent the six HXR peaks of this flare. The right panel shows the
	EUV/FSI 174 {\AA} image overlapped with STIX HXR CLEAN contour maps
	(50\%, 70\%, and 90\% of the maximum). The coronal source region and two foot-points of this
	flare are clearly visible in this image. The blue dotted arc of
	length $2L$ represent the full loop length. Therefore, $L$ is the
	half-loop length of this flare. The dimensions of the coronal source
	region with 50\% contour, marked $a$ and $b$, are used to
	estimate the area of the contour. This area is used to
	determine the volume (V) of the coronal source.}
\label{Fig: Fl20220328_ltc}
\end{figure}

\begin{figure}
\includegraphics[width=0.52\textwidth, height=0.39\textwidth]{f_20220328_lightcurve_n.pdf}
{\tikz\node[coordinate](start1){};}
\includegraphics[width=0.49\textwidth]{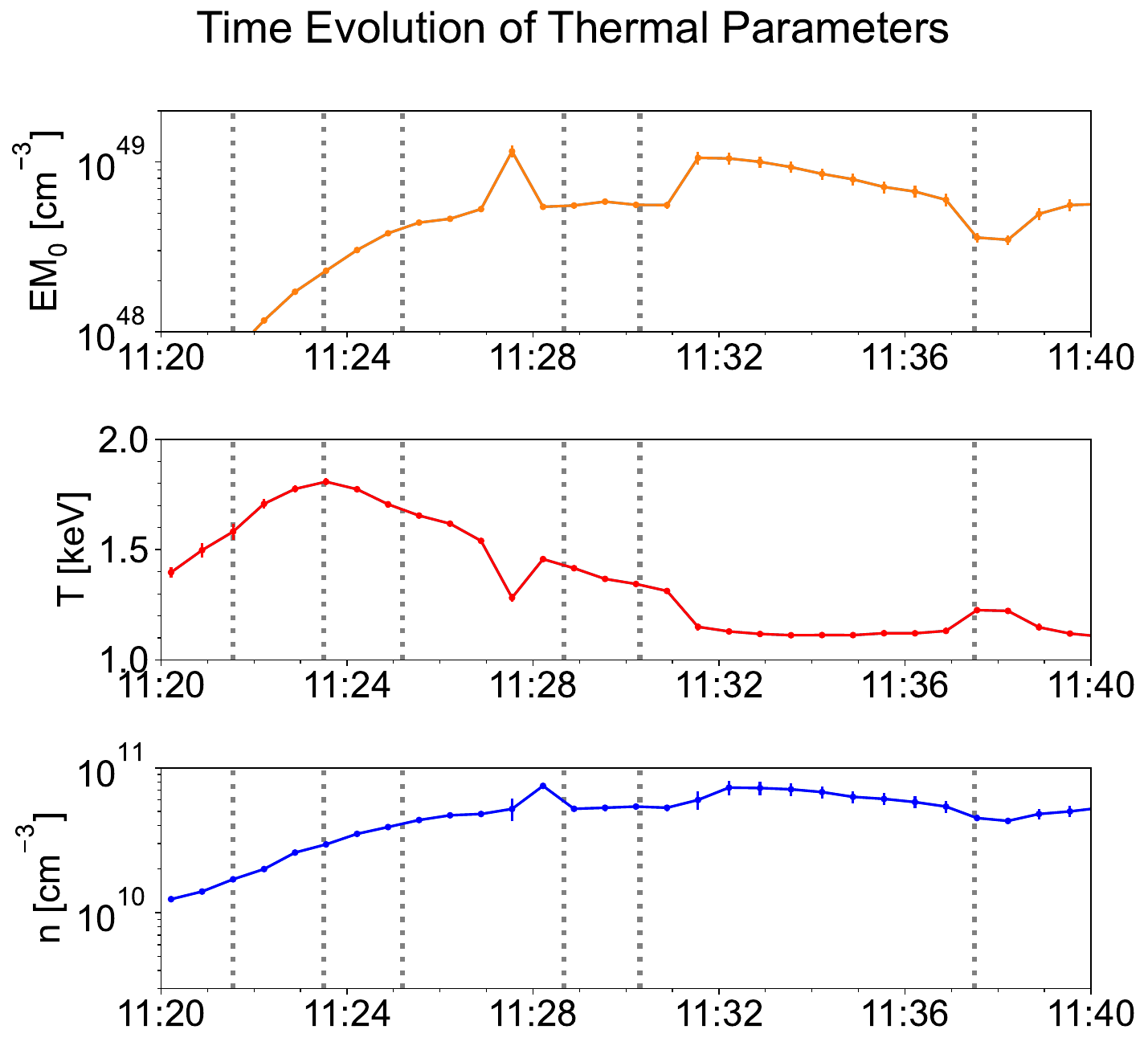}
\includegraphics[width=0.5\textwidth]{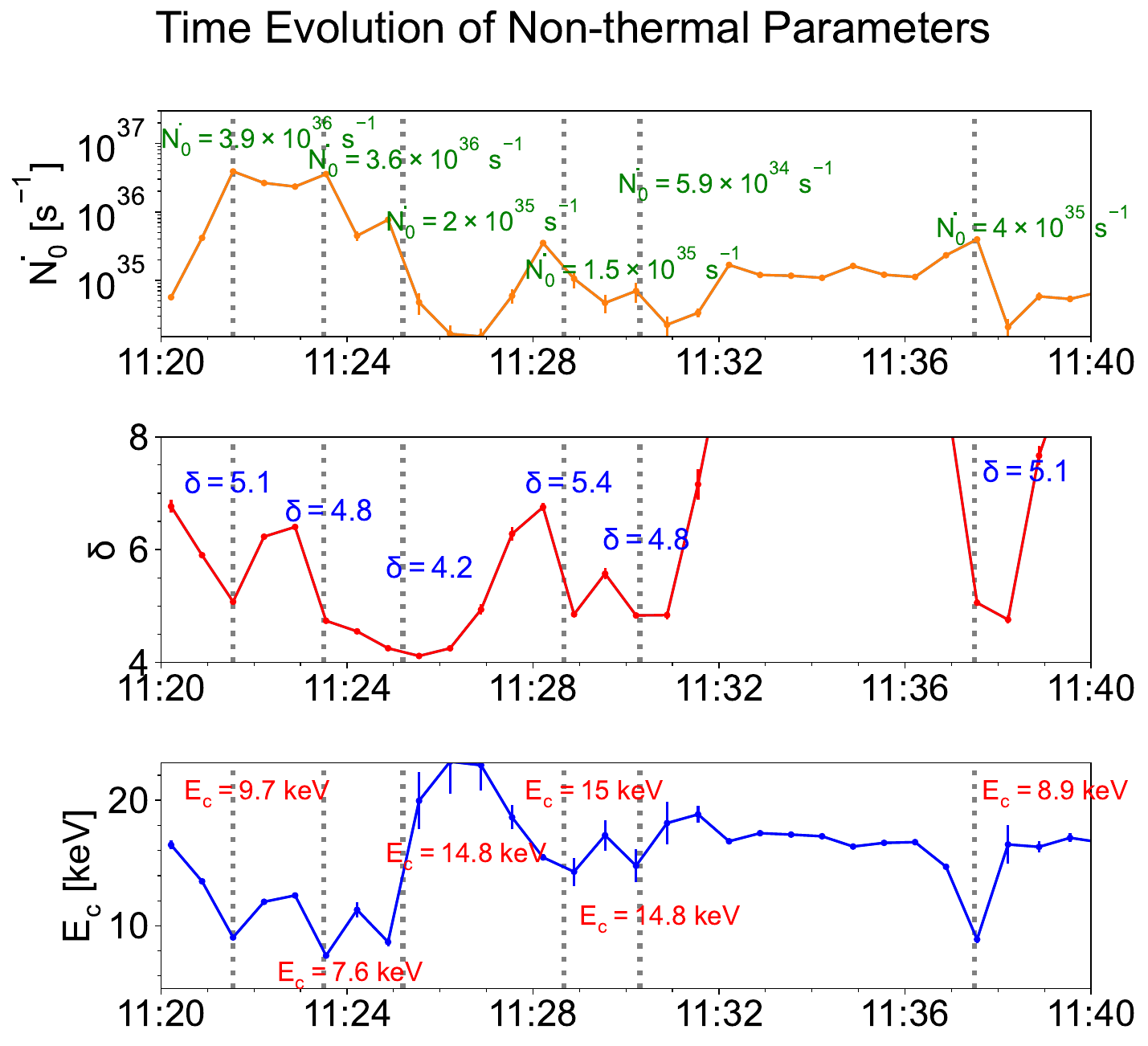}
{\tikz\node[coordinate](end1){};}
\includegraphics[width=0.49\textwidth]{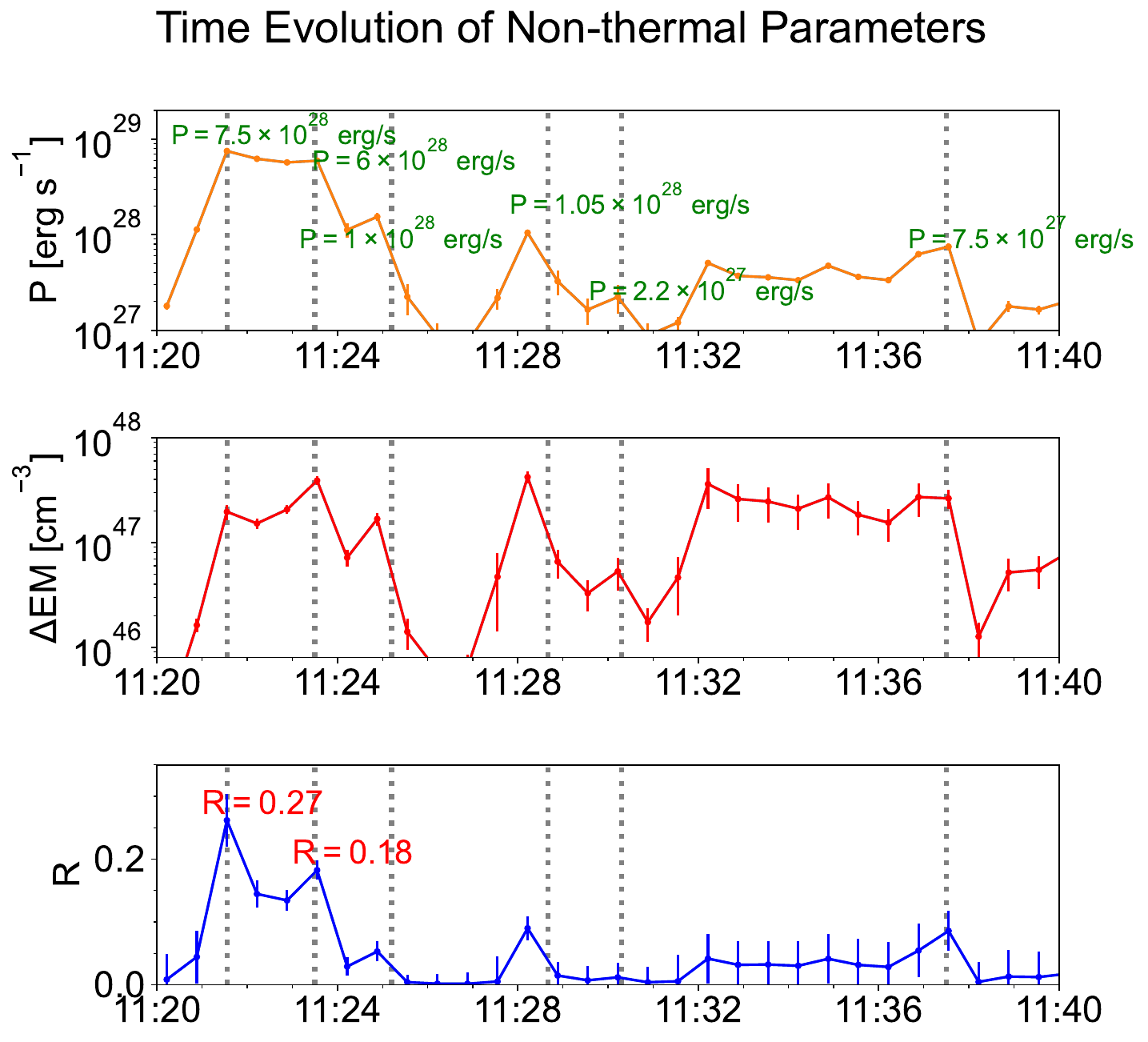}
\caption{The top left panel shows the light curve of the STIX flare
March 28, 2022 as a reference. The region between two black solid vertical lines
represents the full window where we perform the time evolution for this
flare. The top right panel shows the temporal evolution of thermal
parameters derived by fitting the $f_{\rm vth}$ function in the
pre-burst X-ray spectrum. The thermal parameters include: thermal
emission measure ($EM_0$, cm$^{-3}$), plasma temperature ($T$, keV), and
plasma number density ($n$, cm$^{-3}$). The bottom two panels display
the time evolution of nonthermal parameters determined using warm-target
model to the X-ray spectrum of the burst. These include: the total rate of injected
electrons ($\dot{N_0}$, s$^{-1}$), power law index ($\delta$),
low-energy cut-off ($E_c$, keV), total power of nonthermal electrons
($P$, erg s$^{-1}$), excess thermal emission measure from nonthermal
electrons ($\Delta EM$, cm$^{-3}$), and the ratio between $\Delta EM$
and total emission measure ($EM_0 + \Delta EM$), $R$. The six dotted grey vertical lines represent the six HXR peaks of this flare.}

\begin{tikzpicture}[overlay, remember picture, -latex, color=blue!55!red, yshift=150ex, shorten >=0pt, shorten <=0pt, line width=0.03cm]
  \path[-] ($(start1)+(-8.2cm,-0.94cm)$) edge [out=37, in=218]
  ($(end1)+(-5.15cm,9.15cm)$);
\end{tikzpicture}
\begin{tikzpicture}[overlay, remember picture, -latex, color=blue!55!red, yshift=150ex, shorten >=0pt, shorten <=0pt, line width=0.03cm]
  \path[-] ($(start1)+(-0.82cm,-0.94cm)$) edge [out=115, in=295]
  ($(end1)+(-1.22cm,9.15cm)$);
\end{tikzpicture}

\label{Fig: Flare_March28}
\end{figure}

\subsection{Flare: March 28, 2022} \label{S: March28}

The third event in our list is the GOES~M4.0~class March 28, 2022 flare.
The X-ray signatures of this flare are observed by the STIX instrument
\citep{2020krucker} onboard the SolO spacecraft \citep{2020muller}. This
flare is located in the eastern limb with respect to the SolO
spacecraft. We note that a time difference of 335.7 s is included for
this flare during the observation (see discussion by \citet{2024ApJ...964..145J,2024Yingjie}). The temporal evolution of energy
release and transport processes associated with this flare using HXR and EUV observations has been studied in detail by
\cite{2023purkhart}. For this event, the bottom pixels of the instrument
have significantly higher counts than the top pixels
\citep{2024Yingjie}. Therefore, we are using only the bottom pixels for
the current analysis. The light curve with the energy windows: 4-10,
10-12, 12-25, 25-50, and 50-84 keV, is shown in the left panel of
Figure~\ref{Fig: Fl20220328_ltc}. The temporal region (11:20:10 UT to
11:40:10 UT) between two vertical black lines represents the time window
of our interest, while the region in the blue displays the chosen
background for the current analysis. The six gray dotted vertical lines
within the chosen time window (see the left panel of Figure~\ref{Fig:
Fl20220328_ltc}) show the six HXR peaks of this flare. The times
corresponding to these six peaks are: 11:21:33 UT, 11:23:25 UT, 11:25:12
UT, 11:28:40 UT, 11:30:27 UT, and 11:37:46 UT, respectively. The Full
Sun Imager (FSI) which is associated with the Extreme Ultraviolet Imager
(EUI, \cite{2020Rochus}) onboard the SolO provides us with an image of
this flare. This FSI/EUI image of 174 {\AA}, overlapped with the STIX
X-ray contours, is shown in the right panel of Figure~\ref{Fig:
Fl20220328_ltc}. We use the CLEAN contours of 50\%, 70\%, and 90\% of
the maximum for the energy ranges of 8-12 keV, and 25-50 keV, to probe
the coronal loop-top source (red contours) and foot points (green
contours) of this flare, respectively. We use the same procedure as
described for the flares previously analyzed (see \S~\ref{S: Feb24} and
\S~\ref{S: May15}) to estimate the $L$ and $V$ of this flare. Further,
we also find $L = 77 \arcsec \approx 18.6$ Mm (see the blue dotted
semicircular arc marked `2L' in the right panel of Figure~\ref{Fig:
Fl20220328_ltc}), which is in agreement with the value of $L$ estimated
by \cite{2024Yingjie}. By measuring the dimensions of $a$ (the solid black
line marked `$a$' in the right panel of Figure~\ref{Fig:
Fl20220328_ltc}) to be $45 \arcsec \approx 10.8$ Mm, and $b$ (the solid
cyan line marked `$b$' in the right panel of Figure~\ref{Fig:
Fl20220328_ltc}) to be $90 \arcsec \approx 21.7$ Mm, we estimate the
area of the coronal source as $A = 9.2 \times 10^{17}$ cm$^2$. The
source volume is therefore $V \approx 2\times 10^{27}$ cm$^{3}$.
We assume that $L$ and $V$ do not vary with the time evolution of this
flare. Like the two flares discussed previously, the time window of our
interest for this flare (between 11:20:10 UT and 11:40:10 UT) is split
into equal intervals each with 40 s. Each 40 s interval act as the
pre-burst interval for the immediately next interval. We choose 40 s
interval because it agrees with the average diffusion time of the
thermal electrons ($\approx$ 52 s) in this flare. By fitting $f_{\rm vth}
+ f_{\rm thick2}$ to the X-ray spectra of each of the pre-burst
intervals for the energy range 4-80 keV, we determine the time profile
of $EM_0$ and $T$. Using $EM_0$ and $V$, we estimate $n$. The time profiles of these thermal parameters are shown
in the top-right panel of Figure~\ref{Fig: Flare_March28}.

\begin{figure}
	\centering
	\includegraphics[width=0.6\textwidth]{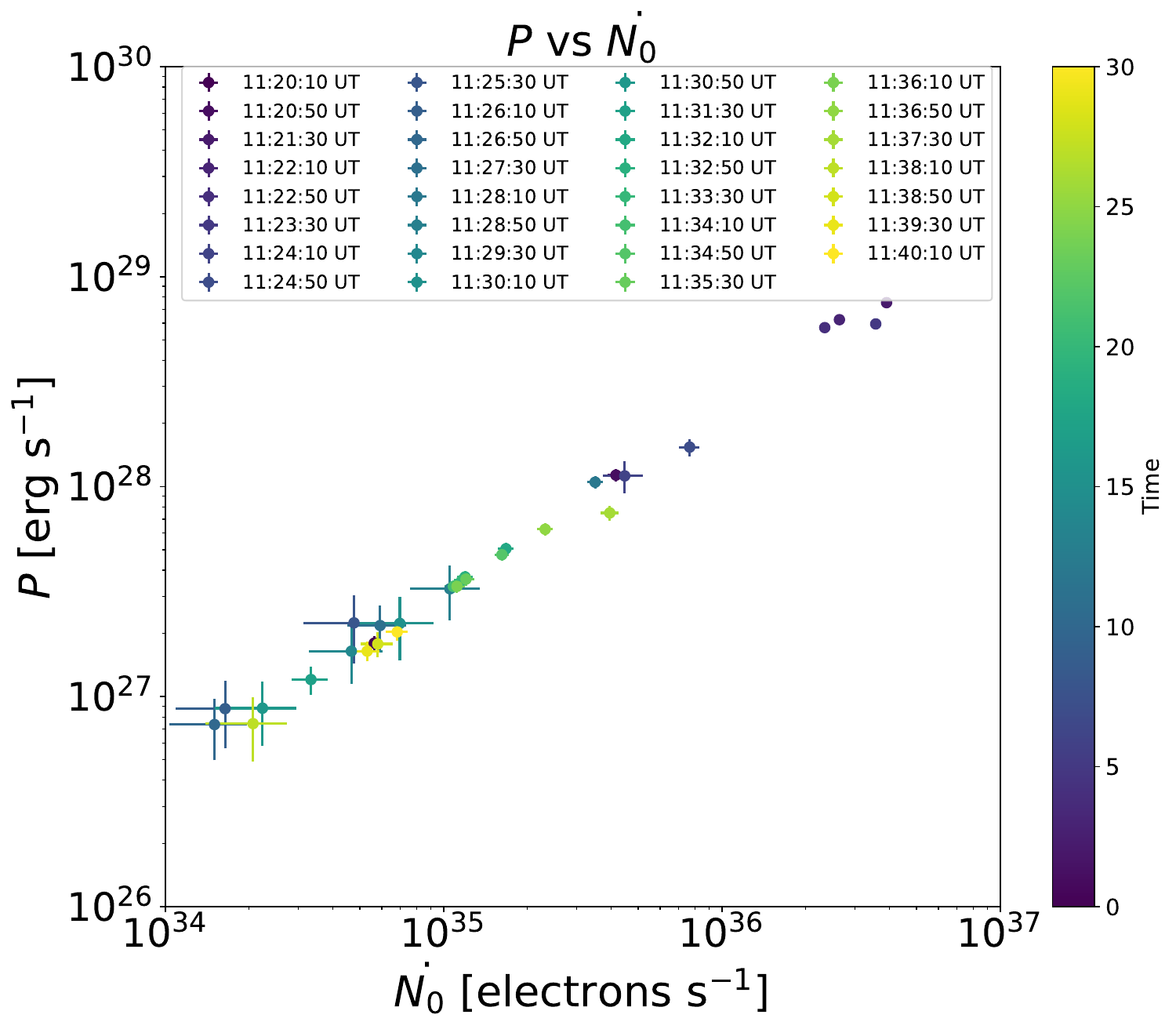}
	\caption{The power of nonthermal electrons ($P$) vs the total rate
	of nonthermal electrons ($\dot{N_0}$) from the Hard X-ray fits for
	the flare March 28, 2022. The legend of this plot shows the selected intervals of this flare.}
\label{Fig: F3_linear}
\end{figure}

\begin{figure}
	\centering
	\includegraphics[width=0.55\textwidth]{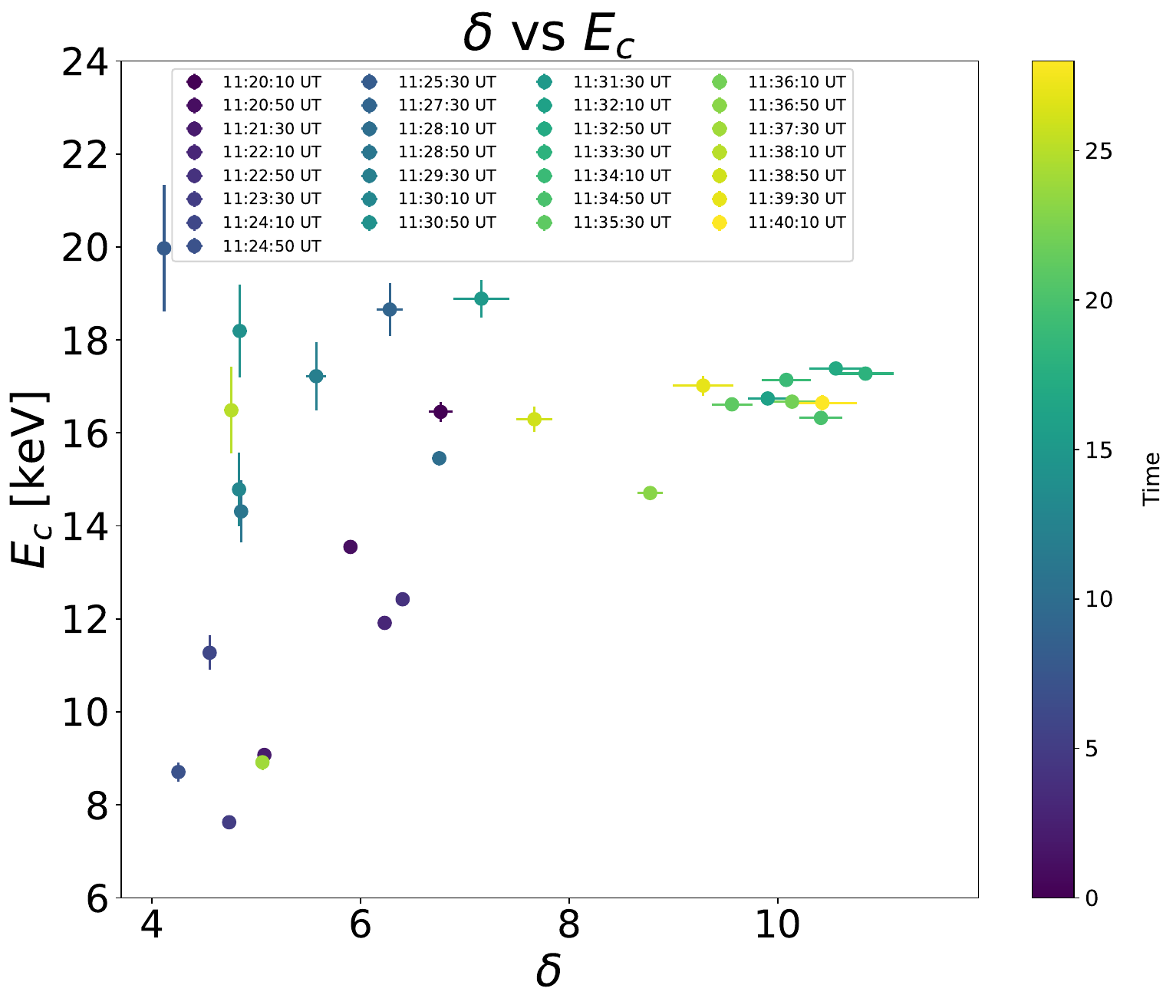}

	\caption{The low-energy cut off of nonthermal electron distribution ($E_{\rm c}$) vs the spectral index ($\delta$) from the Hard X-ray fits for
	the flare March 28, 2022. The legend describes the chosen intervals for this flare. The Pearson's correlation coefficient ($r$) between these two parameters is 0.51 with p-value 0.01.}
\label{Fig: F3_corr}
\end{figure}

Similar to the events analyzed previously, we fix the thermal parameters
while fitting the burst intervals using $f_{\rm vth} + f_{\rm
thick-warm}$ in the energy range 4-80 keV. We note negligible pile up
and albedo corrections for this flare. The best-fit results using the
warm-target model provides us with the temporal evolution of the
parameters associated with nonthermal electrons. The time evolutions of
$\dot{N_0}$, $\delta$, and $E_{\rm c}$ are displayed in the three plots
of the bottom-left panel of Figure~\ref{Fig: Flare_March28}. We note the
soft-hard-soft characteristics is clearly visible around each of the HXR
peaks in the time profile of $\delta$. The two higher peaks of
$\dot{N_0}$ coincide with the first and second HXR peaks with their
magnitudes $(3.9 \pm 0.1) \times 10^{36}$ s$^{-1}$, and $(3.6 \pm 0.1)
\times 10^{36}$ s$^{-1}$, respectively. We also note two smaller peaks
of $\dot{N_0}$ ($\dot{N_0} = (5.9 \pm 0.6) \times 10^{34}$ s$^{-1}$ and $
\dot{N_0} = (4.0 \pm 0.2) \times 10^{35}$ s$^{-1}$) at the positions of the fifth
and sixth HXR peaks. We find that the two peaks of $\dot{N_0}$ immediately
preceding the third and fourth HXR peaks. 

The time profile of $E_{\rm c}$ shows a high-low-high trend around all
the HXR peaks of this flare. The magnitudes of $E_{\rm c}$ at these six
HXR peaks are $9.70 \pm 0.04$ keV, $7.60 \pm 0.08$ keV, $14.8$ keV,
$15.1$ keV, $14.8 \pm 0.1$ keV, and $8.92 \pm 0.08$ keV, respectively.
These local minima of $E_{\rm c}$ coincide with the local maxima of
$\dot{N_0}$ and local minima of $\delta$ throughout the evolution of the
flare. The third HXR peak where $\delta$ (4.2) is close to minimum
coincides with a descending profile of $\dot{N_0}$ and an ascending
profile of $E_{\rm c}$. 
Similar to the previous two flares, we note that a lower magnitude in
$E_{\rm c}$ corresponds to a higher value in the injected electron rate
and a hardness in the accelerated electron spectrum. 
We note that $r$ between the time profiles of $\delta$ and $E_{\rm c}$ is 0.51 with p-value 0.01 (see Fig~\ref{Fig: F3_corr}).\\
The magnitudes of $P$ for each of the six HXR peaks for this flare are
described in the top plot of the bottom-right panel of Figure~\ref{Fig:
Flare_March28}. The first HXR peak (11:21:33 UT) is associated with the
maximum power with magnitude $(7.5 \pm 0.3) \times 10^{28}$ erg
s$^{-1}$. The second HXR peak is associated with the second-largest
value of power ($P = (6.0 \pm 0.4) \times 10^{28}$ erg s$^{-1}$). The
two smaller peaks of $P$ ($P = (2.2 \pm 0.3) \times 10^{27}$ erg
s$^{-1}$ and $P = (7.5 \pm 0.4) \times 10^{27}$ erg s$^{-1}$) coincide
with the fifth and sixth HXR peaks. We note that the other two peaks of
total power immediately precedes the third and fourth HXR peaks.

Similarly to the flares analyzed previously, the time profile of $P$ shows a linear
relationship with the time evolution of $\dot{N_0}$ on the logarithmic scale.
Finally, we seek to understand how the $\Delta EM$ contributes to
the total emission measure ($EM$) throughout this flare. This is shown
in the time profile of $R$ in the bottom plot of the bottom-right panel
of Figure~\ref{Fig: Flare_March28}. It shows that the first peak of $R$ (with $R
= 0.27 \pm 0.12$) coincides with the first HXR peak of this flare. This
is the maximum value of $R$ throughout this flare. At the second HXR
peak, the value of $R$ is found to be $0.18 \pm 0.07$. This suggests
that there is a $27 \pm 12\%$, and $18 \pm 7\%$ contribution of $\Delta
EM$ to $EM$ at the first two HXR peaks, respectively. However, the contribution is below 10\% at the other HXR peaks.

\section{Summary} 
\label{S: summary}

In this study, the time evolution of three well-observed flares is
analyzed using HXR observations and the warm-target model. All flares are
close to the limb, allowing insights into the height structure of the
flaring regions: showing well-defined coronal sources and HXR
footpoints. Using HXR spectroscopy with warm-target model, we estimate
the properties of the nonthermal electron distributions by inferring the
low-energy cut-off of injected electron distribution ($E_{\rm c}$) and the
total power of nonthermal electrons ($P$). These parameters are
normally poorly constrained or unknown within the cold-target model.

The time evolution of $E_{\rm c}$ shows a high-low-high trend around the
HXR peaks. This pattern resembles the soft-hard-soft (high-low-high)
variation of the spectral index $\delta$, often reported in solar flares. Such resemblance can also be seen from the estimated Pearson's correlation coefficient ($r \approx 0.4-0.5$) between these two parameters.
The time profile of $\dot{N_0}$ shows an opposite behavior
(low-high-low) around the HXR peaks, suggesting that a lower value of $E_{\rm
c}$ typically coincides with the spectral hardness around the HXR peaks and
a higher rate of electron acceleration. At the same time, the observed
variations of $E_c$ near the flare peak are rather modest, intriguingly
resembling a the spectrum of acceleration electrons in many stochastic
simulations
\citep[][]{1996ApJ...461..445M,2011ApJ...729..101G,2021PhRvL.126m5101A,2025ApJ...983...58P}.
The time variations of $\dot{N_0}$ are evident in both the observations and simulations.

The total power of nonthermal electrons ($P$) depends on $\dot{N_0}$,
$\delta$, and $E_{\rm c}$. We find that the time evolution of $P$ is
linearly varying with the time profile of $\dot{N_0}$ on the logarithmic scale.
Although $P$ crucially depends on $E_{\rm c}$ and a small change in
$E_{\rm c}$ can significantly change the magnitude of $P$, the time
profile of $P$ predominantly follows the time profile of $\dot{N_0}$.
This linear relationship on the logarithmic scale is also evident from the plots
between the temporal evolutions of $P$ and $\dot{N_0}$ (see the
Figs~\ref{Fig: Fl20110224_linear},\ref{Fig: F2_linear}, and \ref{Fig:
F3_linear}). These outcomes show that the total power of nonthermal
electrons maximizes when the injected electron rate reaches its peak
value and $E_{\rm c}$ reaches a local minimum and electron spectrum
becomes harder. It is likely that the electrons are accelerated at a
lower energy during the peak of the flare. In other words, around the
HXR peak of a flare, there could be an increase of the acceleration
efficiency of electrons. We also note that for all three studied flares, the total nonthermal energy derived from the power for a specific chosen interval is $\approx 10^{30}$ erg and always lower than the corresponding magnitude of bolometric energy with respect to the GOES peak flux as described in Figure 1 of \cite{2020warmuth}.

Finally, we ask 
how the excess thermal emission measure ($\Delta EM$) coming from the
injected electrons compares with the total emission measure ($EM = EM_0
+ \Delta EM$) along the time evolution of flares. For that, the time
profile of $R$ (see Equation~\ref{eq: R}) is investigated. A large $R$
suggests a large fraction of flare-accelerated electrons being
thermalized in the hot coronal part of the flare loop due to collisions
with the ambient plasma. We find that for the two limb flares (February
24, 2011 and March 28, 2022), $R$ shows its maximum at the first HXR
peak of these flares and the magnitude of $R$ then gradually decays
after that. For the flare May 15, 2013, this trend is less prominent as
it does not have multiple HXR impulses like the other two flares.
However, we find that at 01:39:15, $R$ becomes maximum with a magnitude of
$0.25 \pm 0.09$, suggesting a maximum of $25 \pm 9\%$ of the $\Delta EM$
to the total emission measure in this flare. We note that $\Delta EM
\propto \dot{N_0}$ (Eq~\ref{eq: delta_EM}) and $\dot{N_0}$ peaks around
the HXR impulses. $\Delta EM$ is also an estimate of the additional
contribution to the overall inferred soft X-ray (SXR) emission measure
that results from the thermalization of accelerated electrons. The time
evolution of thermal emission measure ($EM_0$) follows that of the
overall SXR impulses. As the first HXR pulse generally peaks before the
SXR maximum, $R$ therefore implies how much the SXR signature that is
coming from the injected electrons, contributes to the overall SXR
signature of the ambient plasma. At the first HXR peak, due to the high
rate of injected electrons, more injected electrons are getting
thermalized due to collisions with the ambient plasma, suggesting an
enhancement in $R$. As time proceeds, $\dot{N_0}$ drops and the ambient
plasma density increases, causing an increase in $EM_0$. This overshadows
the thermal contribution arising from the thermalization of the injected
electrons. Therefore, $R$ decays gradually. This study highlights that
the thermal contribution from the injected electrons is crucial to
characterize the overall total emission measure in the impulsive phase
of a flare that is often interpreted as the result of chromospheric
evaporation
\citep[e.g.][]{1982SoPh...78..107A,1995ApJ...448..915L,2005ApJ...618L.157W}. Since the return current is stronger for stronger direct beam 
and the electron spectrum could be flatter at low energies. 
The recent simulations by \cite{2024alaoui} performed 
for a wide range of solar flare parameters 
provide a way to estimate this effect.

\begin{acknowledgments}
	The work is supported via the STFC/UKRI grants ST/T000422/1 and
    ST/Y001834/1. This research has made use of the Astrophysics Data
    System, funded by NASA under Cooperative Agreement 80NSSC21M00561.
    EPK acknowledges support from the International Space Science
    Institute, Switzerland for the solar flare team
    \href{https://teams.issibern.ch/solarflarexray/}{https://teams.issibern.ch/solarflarexray/}. This paper has benefited from helpful inputs from the anonymous reviewer. The authors acknowledge support from the University of Glasgow to meet the Open Access Publication of this paper.
\end{acknowledgments}
	
	\bibliography{ms}{}
	\bibliographystyle{aasjournal}

\end{document}